\documentclass{article}

\usepackage{arxiv}
\usepackage[square,numbers]{natbib}
\usepackage[utf8]{inputenc} 
\usepackage[T1]{fontenc}    
\usepackage{hyperref}       
\usepackage{url}            
\usepackage{booktabs}       
\usepackage{amsfonts}       
\usepackage{nicefrac}       
\usepackage{microtype}      

\usepackage{lipsum}
\usepackage{enumitem}
\usepackage{amsmath}
\usepackage{graphicx}
\usepackage{subcaption}     
\usepackage{wrapfig}

\usepackage{xcolor}

\usepackage[title]{appendix}

\newcommand{\floor}[1]{\left\lfloor #1 \right\rfloor}

\newcommand{\ceiling}[1]{\left\lceil #1 \right\rceil}

\newtheorem{theorem}{Theorem}
\newtheorem{lemma}{Lemma}

\title{Revisiting Consistent Hashing with Bounded Loads}

\author{
  John Chen\\
  Department of Computer Science\\
  Rice University\\
  Houston, Texas \\
  \texttt{johnchen@rice.edu} \\
   \And
  Ben Coleman\\
  Department of Computer Science\\
  Rice University\\
  Houston, Texas \\
  \texttt{Ben.Coleman@rice.edu} \\
   \And
 Anshumali Shrivastava \\
  Department of Computer Science\\
  Rice University\\
  Houston, Texas \\
  \texttt{anshumali@rice.edu} \\
}

\begin{document}
\maketitle

\begin{abstract}
Dynamic load balancing lies at the heart of distributed caching. Here, the goal is to assign objects (load) to servers (computing nodes) in a way that provides load balancing while at the same time dynamically adjusts to the addition or removal of servers. One essential requirement is that the addition or removal of small servers should not require us to recompute the complete assignment.  A popular and widely adopted solution is the two-decade-old Consistent Hashing (CH) \cite{karg1997og}. Recently, an elegant extension was provided to account for server bounds \cite{vaha2016bounded}. In this paper, we identify that existing methodologies for CH and its variants suffer from cascaded overflow, leading to poor load balancing. This cascading effect leads to decreasing performance of the hashing procedure with increasing load. To overcome the cascading effect, we propose a simple solution to CH based on recent advances in fast minwise hashing. We show, both theoretically and empirically, that our proposed solution is significantly superior for load balancing and is optimal in many senses. On the AOL search dataset and Indiana University Clicks dataset with real user activity, our proposed solution reduces cache misses by several magnitudes.
\end{abstract}


\section{Introduction}

Load balancing is critical to achieve low latency with few server failures and cache misses in networks and web services \citep{karg1997og, chord2001, chord2003}. The goal of load balancing is to assign objects (or clients) to servers (computing nodes referred to as bins) so that each bin has roughly the same number of objects. The \textit{load} of a bin is defined as the number of objects in the bin. In practice, objects arrive and leave dynamically due to spikes in popularity or other events. Bins may also be added and removed due to server failures. The holy grail of distributed caching is to balance load evenly with minimal cache misses and server failures. Poor load balancing directly increases latency and cost of the system~\cite{SPOCA}.

Caching servers often use hashing to implement dynamic load assignment. Traditional hashing techniques, which assign objects to bins according to fixed or pre-sampled hash codes, are inappropriate because bins are frequently added or removed. Standard hashing and Cuckoo hashing \cite{Cuckoo1, Cuckoo2, Cuckoo3, Cuckoo4} are inefficient because they reassign all objects when a bin is added or removed.

Consistent Hashing (CH) \cite{karg1997og} is a widely adopted solution to this problem. In CH, objects and bins are both hashed to random locations on the unit circle. Objects are initially assigned to the closest bin in the clockwise direction (see Figure~\ref{fig:CH} and Section~\ref{CHsection}). CH is efficient for the dynamic setting because the addition or removal of a bin only affects the objects in the closest clockwise bin. 

In practice, we cannot assign an unlimited number of objects to a bin without crashing the corresponding server. In~\cite{vaha2016bounded}, the authors address the problem by setting a maximum bin capacity $C = \ceiling{(1 + \epsilon)\frac{n}{k}}$, where $n$ objects are assigned to $k$ bins, each with a capacity parameter $\epsilon \geq 0$. Their hashing scheme ensures assigns new objects to the closest non-full bin in the clockwise direction and ensures that the maximum load is bounded by $C$. There are also many hueristics, such as time-based expiry and eviction recommended in ASP.net \cite{cachingASP}, Microsoft \cite{cachingMicrosoft}, Mozilla \cite{cachingMozilla}.



\textbf{Applications:} 
Dynamic load assignment is a fundamental problem with a variety of concrete, practical applications. CH is a core part of Discord's 250 million user chat app \cite{discord}, Amazon's Dynamo storage system \cite{Dynamo} and Apache Cassandra, a distributed database system \cite{Cassandra}. Google cloud and Vimeo video streaming both use CH with load bounds \cite{CHBDblog, vimeo}. CH is also used for information retrieval \cite{CHapplications2}, distributed databases \cite{CHDistributedDatabase1, CHDistributedDatabase2, CHDistributedDatabase3}, and cloud systems \cite{CHCloudSystems1, CHCloudSystems2, CHCloudSystems3}. Furthermore, CH resolves similar load-balancing issues that arise in peer-to-peer systems \cite{peerToPeerSystems, peerToPeerSystems2}, and content-addressable networks \cite{contentAddressableNetworks}. 

\textbf{Our Contributions:} We propose a new dynamic hashing algorithm with superior load balancing behavior. To minimize the risk of overloading a bin, all bins should ideally have approximately the same number of objects at all times. Existing algorithms experience a cascading effect that unevenly loads bins with the clockwise object assignment procedure. 

Our algorithm improves upon the load balancing problem both in theory and practice. In our experiments on real user logs from the AOL search dataset and Indiana University Clicks dataset~\cite{Meiss08WSDM}, the algorithm reduces cache misses by several orders of magnitude. We prove optimality for several criteria and show that the state-of-the-art method stochastically dominates the proposed method. The experiments and theory show that our algorithm provides the most even distribution of bin loads. \vspace{-0.3cm}


\section{Background}
\vspace{-0.1cm}
\subsection{2-Universal Hashing}\vspace{-0.05cm}
A hash function $h_{\mathrm{univ}} : [l] \rightarrow [m]$ is 2-universal if for all $i, j \in [l]$ with $i \neq j$, we have the following property for any $z_1, z_2 \in [m]$, \vspace{-0.2cm}
\begin{equation*}
Pr(h_{\mathrm{univ}}(i) = z_1 \text{ and } h_{\mathrm{univ}}(j) = z_2) = \frac{1}{m^2} \text{ .} \vspace{-0.1cm}
\end{equation*}
\subsection{Consistent Hashing} \label{CHsection}
\begin{figure}[h!]
  \centering \vspace{-0.5cm}
  \begin{subfigure}[b]{0.30\linewidth}
    \includegraphics[width=\linewidth]{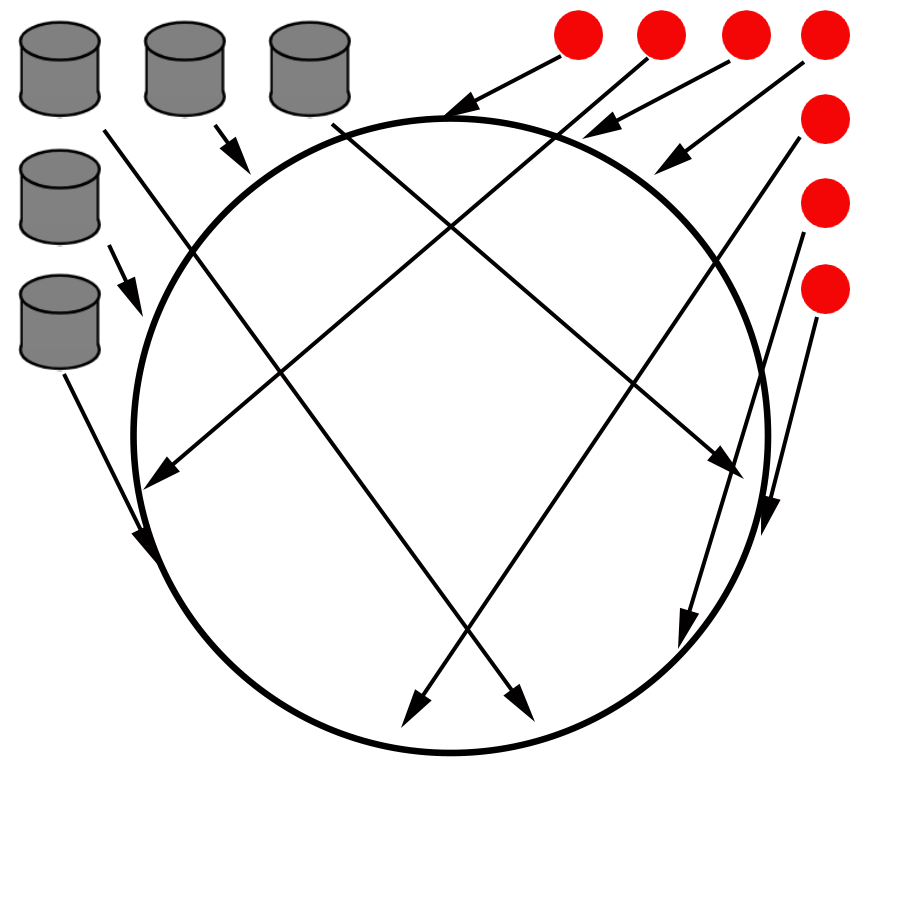}\vspace{-0.3cm}
    \caption{Initial placement. \newline \vspace{0.2cm}} 
    \label{fig:CHa}
  \end{subfigure}
    \begin{subfigure}[b]{0.30\linewidth}
    \includegraphics[width=\linewidth]{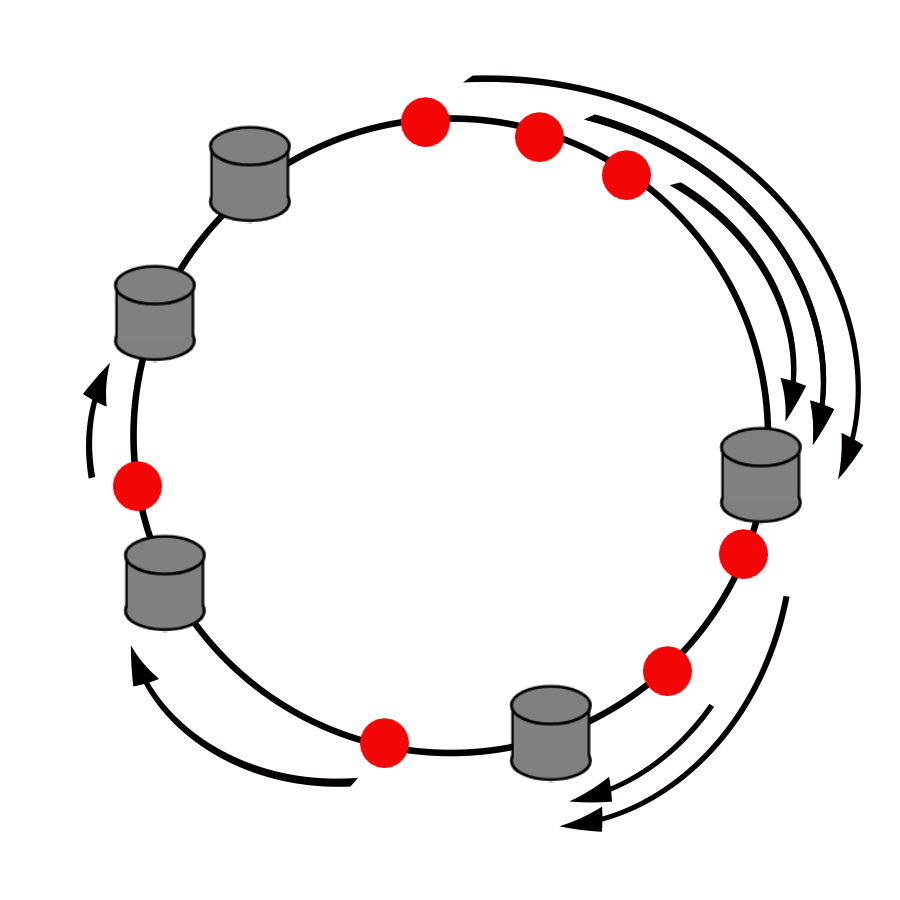}\vspace{-0.3cm}
    \caption{Objects are assigned to the closest bin in clockwise direction. \vspace{0.2cm}}
    \label{fig:CHb}
  \end{subfigure}
    \begin{subfigure}[b]{0.30\linewidth}
    \includegraphics[width=\linewidth]{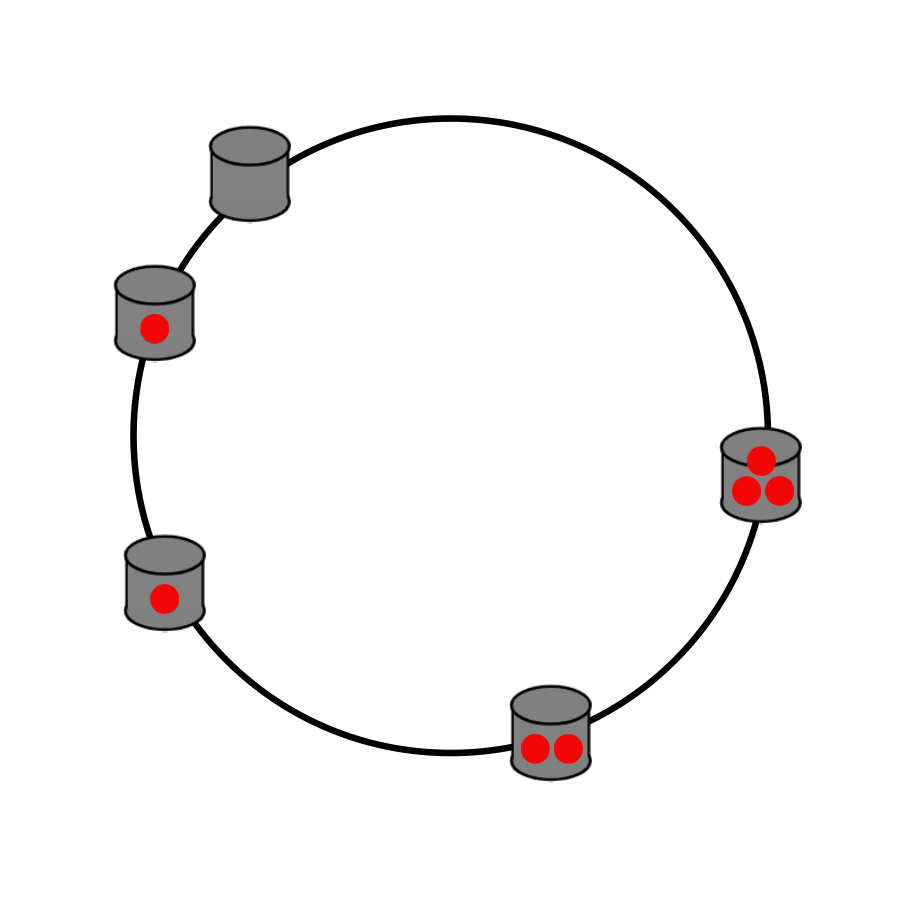} \vspace{-0.3cm}
    \caption{After assignment.\newline \vspace{0.2cm}}
    \label{fig:CHc}
  \end{subfigure}
  \vspace{-0.5cm}
  \caption{Consistent Hashing object and bin assignment. Objects are red. \vspace{0.5cm}}
  \label{fig:CH}
\end{figure}

In the CH scheme, objects and bins are hashed to random locations on the unit circle as shown in Figure \ref{fig:CHa}. Objects are assigned to the closest bin in the clockwise direction, shown in Figure \ref{fig:CHb}, with the final object bin assignment in Figure \ref{fig:CHc}.
 \vspace{-1cm}

When a bin is removed, its objects are deposited into the next closest bin in the clockwise direction the next time they are requested. When a bin is added, it is used to cache incoming objects. Both procedures only reassign objects from one bin, unlike the naive hashing scheme. 
The arc length between a bin and its counter-clockwise neighbor determines the fraction of objects assigned to the bin. In expectation, the arc lengths are all the same because the bins are assigned to the circle via a randomized hash function. With equal arc lengths, each bin has the ideal load of $n/k$. However, CH seldom provides ideal load balancing because the arc lengths have high variance. 



\vspace{-0.3cm}
\subsection{Consistent Hashing with Bounded Loads}

Consistent Hashing with Bounded Loads (CH-BL) was proposed by \cite{vaha2016bounded} to model bins with finite capacity. CH-BL extends CH with a maximum bin capacity $C = \ceiling{(1 + \epsilon)\frac{n}{k}}$. Here, $n$ is the number of objects, $k$ is the number of bins, and $\epsilon \geq 0$ controls the bin capacity.

\begin{wrapfigure}{r}{0.20\linewidth}
  \centering
    \begin{subfigure}[b]{1\linewidth}
    \vspace{-0.5in}
    \includegraphics[width=\linewidth]{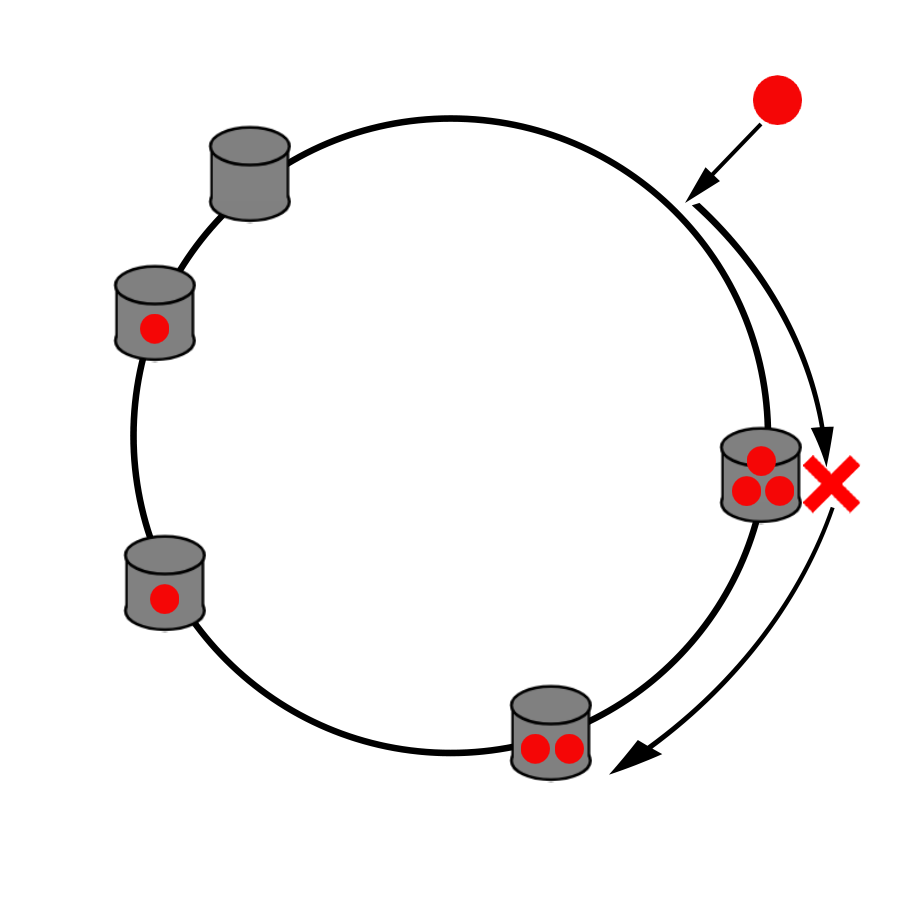}
    \caption{New object arrives.}
  \end{subfigure}
  \caption{CH-BL with bin capacity of 3.\vspace{-0.05cm}}
    \label{fig:CHwithBoundedLoads}
  \centering
    \begin{subfigure}[b]{1\linewidth}
    \includegraphics[width=\linewidth]{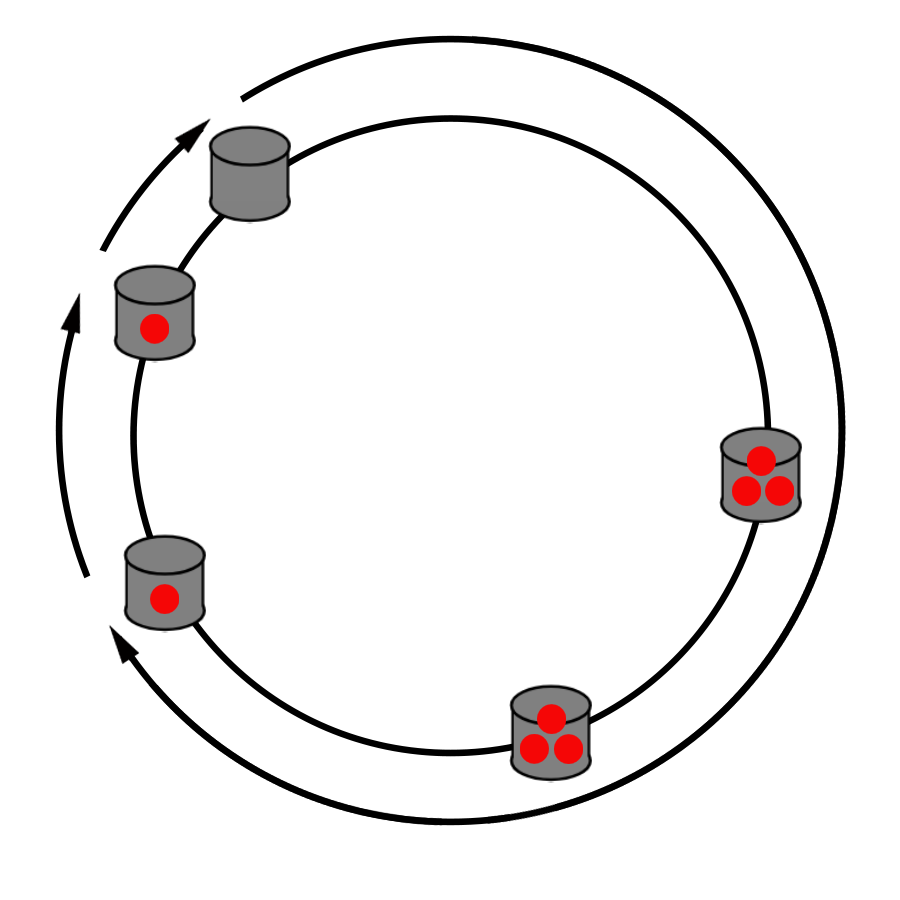}
  \end{subfigure}
  \caption{Effective arclength of each non-full bin. Cascaded overflow of CH-BL with bin capacity of 3.}
  \label{fig:CHwithBoundedLoadsCascadedOverflow} \vspace{-1.5cm}
\end{wrapfigure}

In CH-BL, if an object is about to be assigned to a full bin, it overflows or cascades into the nearest available bin in the clockwise direction. Figure \ref{fig:CHwithBoundedLoads} uses the bin object assignment from Figure \ref{fig:CH} as the initial assignment with a maximum bin capacity of 3. A new object is hashed into the unit circle, but the closest bin in the clockwise direction is unavailable because it is full. Therefore, this object is assigned to the nearest available bin.

On bin removal, CH-BL performs the same reallocation procedure as CH, but with bounded loads. Objects from a deleted bin are cached in the closest available bin in the clockwise direction the next time the object is requested. Bin addition is handled the same as CH. 

\vspace{-0.2cm}
\subsection{Cascaded Overflow of Consistent Hashing and Variants}

CH-BL solves the bin capacity problem but introduces an overflow problem. Recall that the expected number of objects assigned to a particular bin is proportional to the bin's arc length. As bins fill up in CH-BL, the nearest available (non-full) bin has a longer and longer effective arc length. The arc lengths for consecutive full bins add, causing the nearest available bin to fill faster. We call this phenomenon \emph{cascaded overflow}. 

Figure~\ref{fig:CHwithBoundedLoadsCascadedOverflow} shows cascaded overflow for the non-full bins in Figure~\ref{fig:CHwithBoundedLoadsCascadedOverflow} using the final object bin assignment in Figure~\ref{fig:CHwithBoundedLoads} with a maximum bin capacity of 3. One bin now owns roughly 75\% of the arc, so it will fill quickly while other bins are underutilized. The cascading effect creates an avalanche of overflowing bins that progressively cause the next bin to have an even larger arc length. 


Cascaded overflow is a liability in practice because overloaded servers often fail and pass their loads to the nearest clockwise server. Cascaded overflow can trigger an avalanche of server failures as an enormous load bounces around the circle, crashing servers wherever it goes. In severe cases, this can bring down the entire service~\cite{SPOCA}. 

\vspace{-0.2cm}
\subsection{Simple Rehashing}\vspace{-0.2cm}
At first glance, one reasonable approach is to rehash objects that map to a full bin rather than use the nearest clockwise bin. We reassign an object to bin $h(i)$ rather than $i+1$ if bin $i$ is full. However, linear probing with random probes fails because it effectively rearranges the unit circle. Bin $i$ always overflows into $h(i)$, preserving the cascaded overflow effect.  \vspace{-0.2cm}

\section{Random Jump Consistent Hashing}
\vspace{-0.1cm}

Our proposal is motivated by Optimal Densification~\cite{ansh2017OD}, a technique introduced to quickly compute minwise hashes in information retrieval. We break the cascade effect by introducing Random Jumps for Consistent Hashing (RJ-CH). In practice, the segments of the unit circle are mapped to an array. RJ-CH continuously rehashes objects until they reach an index associated with an available bin. Unlike simple rehashing, the RJ-CH hash function takes \textit{two} arguments: the object and the failed attempts to find an available bin. The second argument breaks the cascading effect because it ensures that two objects have a low probability of overflowing to the same location. This probability is $1 / m$, where $m$ is the length of the array. 

\begin{figure}[h!]
  \centering\vspace{-0.6cm}
    \begin{subfigure}[b]{0.28\linewidth}
    \includegraphics[width=\linewidth]{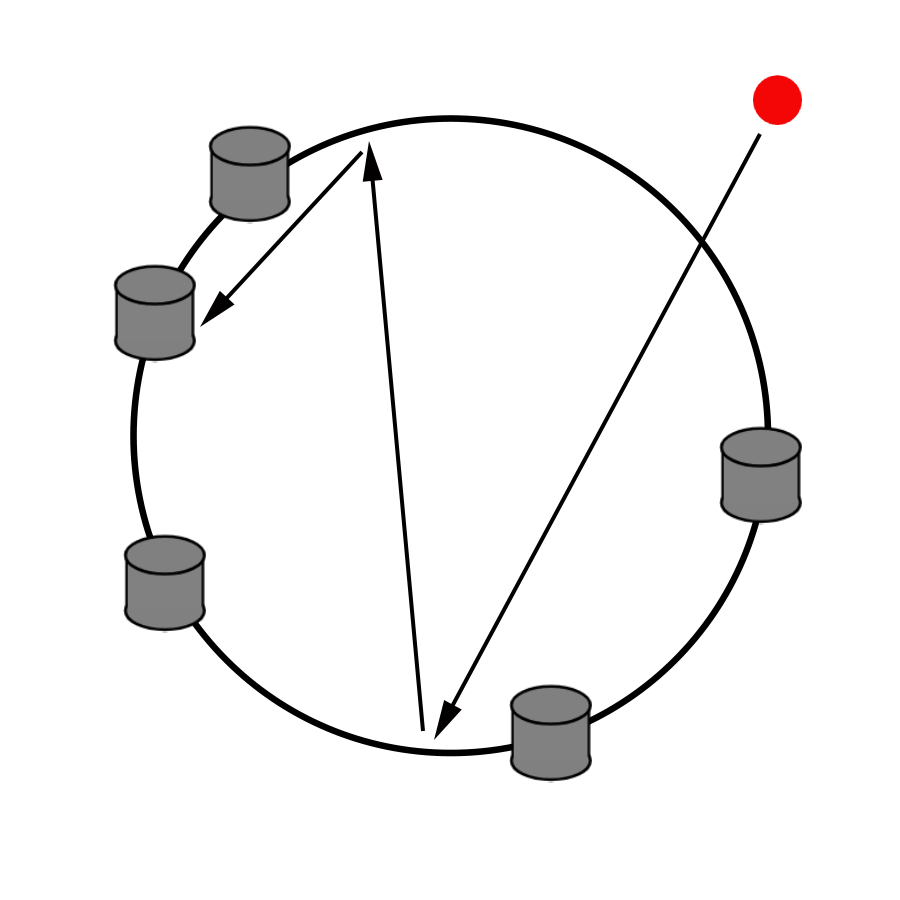} \vspace{-0.6cm}
    \caption{Initial assignment method.\vspace{-0.1cm}}
    \label{fig:CHODa}
  \end{subfigure}
  \begin{subfigure}[b]{0.28\linewidth}
    \includegraphics[width=\linewidth]{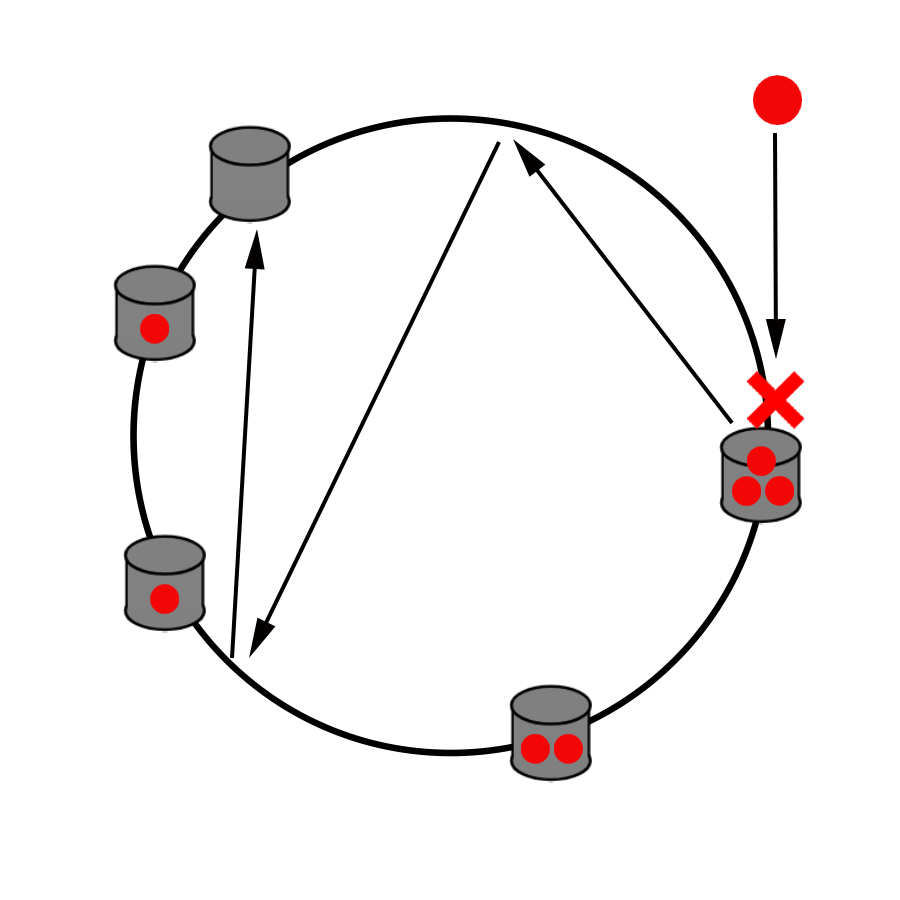}
    \vspace{-0.6cm}
    \caption{New object arrives.\vspace{-0.1cm}}
    \label{fig:CHODb}
  \end{subfigure}
  \begin{subfigure}[b]{0.28\linewidth}
    \includegraphics[width=\linewidth]{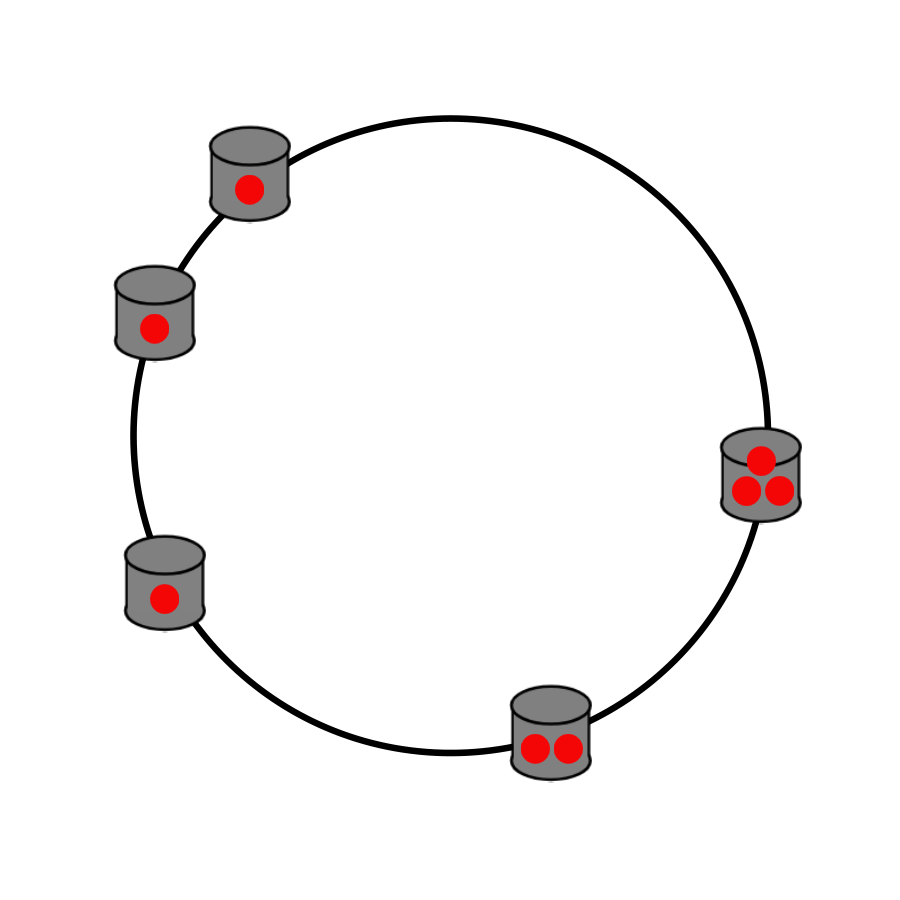} \vspace{-0.6cm}
    \caption{Final assignment.\vspace{-0.1cm}}
    \label{fig:CHODc}
  \end{subfigure}  
  \caption{RJ-CH object and bin assignment with bin capacity of 3.\vspace{0.5cm}}
  \label{fig:CHOD}
\end{figure}

Figure~\ref{fig:CHODa} shows RJ-CH in a situation without full bins, which evolves into Figure \ref{fig:CHODb} when a bin becomes full. RJ-CH prevents cascaded overflow because objects are assigned to any of the available bins with uniform probability by the universal hashing property. RJ-CH cannot be implemented with a dynamically changing array size, but this limitation is is common to RJ-CH, CH-BL and CH. We also note that load balancing methods are usually accompanied by hueristics like time-based expiry and eviction of stale objects~\cite{cachingASP,cachingMicrosoft,cachingMozilla} to evict duplicates and unused objects. Objects are commonly deleted when they are unused for some time. Many implementations, such as \cite{cachingASP}, impose stringent eviction criteria. It is also common practice to wipe the cache of a failed server and repopulate the cache as needed when the server is back online. RJ-CH is compatible with all such techniques, since deleting an element simply frees space in the bin. 

 \vspace{-0.3cm}
\subsection{Discussion: object removal, bin removal and bin addition schemes}
\label{appendix:objectRemoval}\vspace{-0.2cm}


When a bin is added, we may encounter a situation where an object is cached in the new bin while also existing somewhere else in the array. In practice, this is not a problem because the system will no longer request the duplicate and it will eventually be evicted by its bin. When a bin is removed, its objects will be cached in the available bin chosen by RJ-CH the next time the objects are requested. \vspace{-0.3cm}

\section{Theoretical Analysis}
\label{sec:headings}\vspace{-0.2cm}
In this section, we prove that the bin load under CH-BL stochastically dominates that of RJ-CH, showing that RJ-CH has lower bin load variance, fewer full bins and other desirable properties. In addition, the variance of CH-BL increases exponentially as bins become full. RJ-CH also achieves an algorithmic improvement over CH-BL for object insertion.

\vspace{-0.1cm}
\subsection{Bin load following CH-BL stochastically dominates RJ-CH}
\vspace{-0.1cm}


When reassignments are necessary, RJ-CH reassigns objects uniformly to the available bins, while CH-BL reassigns objects to the nearest clockwise bin. Even before a CH-BL bin fills, the object assignment probabilities are unequal as discussed in sections \ref{objectsTillFirstOverflow} and \ref{varianceBeforeFirstOverflow}. Here, we assume that the CH-BL assignment probabilities are initially equal, corresponding to optimal initial bin placements. Let $n$ objects be assigned to $k$ bins with a maximum capacity $C$. Our main theoretical result is as follows. It shows that RJ-CH is superior to CH-BL in terms of smaller variance of the number of objects in each bin, and in terms of the mean number of full bins. Detailed proofs are provided in the Appendix. The main result is as follows: \vspace{-0.2cm}
\begin{theorem} \label{mainTheorem}
Let $X^{(CH-BL)}_i$  ( $X^{(RJ-CH)}_i$ )  denote the number of objects in bin $i$ when placing $n$ objects into a ring of $k$ bins with CH-BL or RJ-CH. Then,\vspace{-0.1cm}
\begin{equation}
var( X^{(RJ-CH)}_i )  \leq   var( X^{(CH-BL)}_i ),
\hbox{for $i=1,..., k$}.
\end{equation}
Moreover, 
\begin{equation} \label{corollary3}
E(L^{(RJ-CH)}) \leq E(L^{(CH-BL)}), 
\end{equation}
where  $L^{(CH-BL)}$ ($L^{(RJ-CH)}$)  is the number of full bins following the CH-BL (RJ-CH) method.
\end{theorem} \vspace{-0.2cm}
Theorem \ref{mainTheorem} is a straightforward special case of the below Theorem \ref{theorem1}. Proof is given in Appendix \ref{corollaryAppendix}. \vspace{-0.2cm}


\begin{theorem}
\label{theorem1}
Let $f(\cdot)$ be a convex function defined on $\{0, 1, ..., C\}$. Then,\vspace{-0.1cm}
\begin{equation} \label{theorem1.1}
\sum_{i=1}^k E[  f( X^{(RJ-CH)}_k ) ] \leq \sum_{i=1}^k E[   f(  X^{(CH-BL)}_i ) ].
\end{equation}
And the symmetry implies
\begin{equation}  \label{theorem1.2}
E[  f( X^{(RJ-CH)}_i ) ] \leq  E[  f(  X^{(CH-BL)}_i ) ], \qquad \hbox{for $i=1,..., k$}.
\end{equation}
\end{theorem}



\vspace{-0.1cm}
The main idea of the proof of Theorem \ref{theorem1} is to consider a scheme where the first $j + 1$ objects are assigned using CH-BL and the rest are assigned using RJ-CH. Such a scheme is worse than, stochastically dominates, a scheme where the first $j$ objects are assigned using CH-BL and the rest are assigned using RJ-CH. Only the $j + 1$th object of the two schemes follow a different assignment method. One key difficulty in the analysis lies in the fact that the differing assignment of that $j + 1$th object affects the assignment of the remaining objects. Lemma \ref{lemma1} proves an equivalent assignment method which allows the $j + 1$th object to be assigned last. Therefore, for the two schemes we only need to consider the "badness" of the last object, since all previous $n - 1$ objects are assigned the same way. Lemmas \ref{lemma2}, \ref{lemma3}, \ref{lemma4} give us the assignment probability of that last object and tools to determine the stochastic dominance of the bin load of one scheme over the other. Lemma \ref{lemma5} completes the proof.
\begin{lemma} \label{lemma1}
Suppose bin $i$ already contains $b_i$ objects, with $b_i < C$ for $i=1,..., K$. Distribute $N$ more objects into the $K$ bins in the following scheme indexed by $m$: 
All objects are assigned uniformly to $K$ bins and relocated following RJ-CH, except for the $m$-th object, which is assigned to bin 1, and reassigned following RJ-CH. Then, the final joint distribution of the numbers of objects in the $K$ bins will be the same regardless of the value of $m=1,..., N$.
\end{lemma}
\vspace{-0.1cm}
Consider again the scheme in which the first $j + 1$ objects are assigned following CH-BL and the remaining $n - (j + 1)$ objects are assigned following RJ-CH. The implication of Lemma \ref{lemma1} is given that the $j + 1$th object was assigned to a bin $i$, it can equivalently be assigned as the $n$th object to bin $i$. If bin $i$ is full, then the object is reassigned using RJ-CH.

Denote ${\cal M}(n;p_1,...,p_k)$ the multinomial distribution for the number of objects in $k$ bins when assigning $n$ objects to $k$ bins where each object has probability $p_i$ of being assigned to bin $i$. Let ${\cal M}_C(n;p_1,...,p_k)$ be the constrained multinomial distribution for the number of objects in $k$ bins when assigning $n$ objects to $k$ bins where each object has probability $p_i$ of being assigned to bin $i$ under the condition that each bin has at most $C - 1$ objects. Let $X_i$ be the random number of objects in bin $i$.\vspace{-0.1cm}
\begin{lemma}
\label{lemma2}
If $(X_1,..., X_k)$ $\sim$  ${\cal M}  (n; p_1, ..., p_k)$, the conditional distribution of $(X_{i_1}, ..., X_{i_J})$ subject to $\sum_{j=1}^J X_{i_j} = n^*$ is ${\cal M} (n^*; p_{1}^*, ..., p^*_{J})$ where $p^*_{ j} = p_{i_j}/ \sum_{l=1}^J p_{i_l}$.   Moreover, if $(X_1,..., X_k)$ $\sim$ ${\cal M}_C  (n; p_1, ..., p_k)$, the conditional distribution of $(X_{i_1}, ..., X_{i_J})$ subject to $\sum_{j=1}^J X_{i_j} = n^*$ is  ${\cal M}_C (n^*; p_{1}^*, ..., p^*_{J})$.
\end{lemma}\vspace{-0.1cm}
Lemma \ref{lemma2} can be understood as the distributions describing the results of assigning $n$ objects randomly to $k$ bins.

A random variable $X$ is stochastically smaller than $Y$, denoted as $X \prec Y$,  if $P(X > x) \leq P(Y >x)$ for all $x$, or, equivalently, if $E(g(X)) \leq E(g(Y))$ for any bounded increasing function $g$.
\vspace{-0.1cm}
\begin{lemma}
\label{lemma3}
Let $\Delta_i$, $i=1,..., n$, be independent random binary random variables taking value 1 with probability $p_i$ and taking value 0 with probability $q_i = 1 - p_i$. Assume $p_i \leq 1/2 \leq q_i$. Let $\xi_1 = \sum_{i=1}^n \Delta_i$ and $\xi_2= n- \xi_1 $.
Then,
\begin{equation} \label{lemma3.1}
P(\xi_1=x) \leq P(\xi_2= x )    \,\,\hbox{and}\,\, P(\xi_1=n-x) \geq P(\xi_2=n-x)
\hbox{  for } \,\,  n/2 \leq x \leq n.
\end{equation}\vspace{-0.1cm}
Moreover, 
\begin{equation} \label{lemma3.2}
P(\xi_1=x | \xi_1 < C, \xi_2 < C) \leq   P(\xi_2= x  | \xi_1 < C, \xi_2 < C) ,
\hbox{for } \max(n/2, n - C) \leq x < C.
\end{equation}
Consequently, $\xi_1 \prec \xi_2$ and \vspace{-0.1cm}
\begin{equation} \label{lemma3.3}
\xi_1  \, | \, (\xi_1 < C, \xi_2 < C)  \prec \xi_2 \, | \, (\xi_1 < C, \xi_2 < C).
\end{equation}
\end{lemma}\vspace{-0.3cm}
If we know the assignment probability $p$ of a bin is greater than another, then Lemma \ref{lemma3} can be used to determine the stochastic dominance of the bin load of one bin over the other.
\vspace{-0.1cm}
\begin{lemma}
\label{lemma4}
Place $n$ objects into $k$ bins following CH-BL. Let $X_i$  be the number of objects in bins $i$, and $L_i$ be the length of cluster of full bins to the right of bin $i$, for $i=1, ..., k$. $L_i=0$ if the bin to the right hand side of bin $i$ is non-full. Let $i_1, ..., i_J$ be all the non-full bins. Then, conditioning on $L_{i_j}$, $j=1,..., J$, and $\sum_{j=1}^J x_{i_j} = n^*$, $(X_{i_1}, ..., X_{i_J})$ follows the constrained multinomial distribution, i.e.,
\begin{equation} \label{lemma4.1} \hbox{the conditional distribution of } (X_{i_1}, ..., X_{i_J}) \sim {\cal M}_C ( n^*; p_1^*, ..., p_J^*),
 \end{equation}
 where $p_j^* = (L_{i_j}+1)/k$ for $j=1, ..., J$, and $n^* = n - (k-J) C$. 
\end{lemma}
\vspace{-0.1cm}
Lemma \ref{lemma4} proves that in expectation bins on the left of longer clusters of full bins have more objects.
\vspace{-0.1cm}
\begin{lemma}
\label{lemma5}
Assign $n = m + 1 + (n - (m + 1))$ total objects into $k$ bins in a scheme with following three steps:\vspace{-0.2cm}
\begin{enumerate}
\item[1.]  Assign $m$ objects following CH-BL. Let ${\cal N} = \{ i_1,..., i_J\}$ denote all the non-full bins, with $L_{i_j}$ as the length of the cluster of full bins to the right of bin $i_j$. For notational simplicity, assume $L_{i_1} \leq ... \leq L_{i_J}$.\vspace{-0.2cm}
\item[2.] Assign one object into bins $i_j $ with probability $q_j $, $j=1,..., J$, such that $\sum_{j=1}^J q_j =1 $ and $0\leq q_1 \leq ...  \leq q_J$, and $q_j$ depends on $L_{i_j}$ only.\vspace{-0.2cm}
\item[3.] Assign $n - (m + 1)$ objects into the bins $1, ..., k$ following RJ-CH.\vspace{-0.2cm}
\end{enumerate}
Let $X_1,..., X_k$ be the numbers of objects in bins $1, ..., k$, and let $f(\cdot)$ be any convex function on $\{0, ..., C\}$. 
Then, \vspace{-0.1cm}
\begin{equation} \label{lemma5.1}
\sum_{i=1}^k E [ f(X_i) ] \hbox{  is minimized when the distribution in Step 2 is uniform, i.e., }
 q_1=\cdots = q_J= 1/J .
\end{equation}
\end{lemma}
\vspace{-0.3cm}
{\bf Proof of Theorem \ref{theorem1}} In Lemma 5 if all $q_j$ are equal, Steps 1-3 are the same as assigning the first $m$ objects following CH-BL and rest $n - m$ objects following RJ-CH. If the first $m + 1$ objects are assigned following CH-BL then $q_j \propto L_{i_j}+1$, and the rest $n - (m + 1)$ objects are assigned following RJ-CH. For the first scheme, we denote by $X^{(m)}_1, ..., X^{(m)}_k$ as the final numbers of objects in bins $1,..., k$. With this notation, $X^{(m+1)}_1, ..., X^{(m+1)}_k$ are the final numbers of objects in bins $1, ..., k$ by the latter method. 
Then, Lemma 5 proves that\vspace{-0.1cm}
\begin{equation*}
\sum_{i=1}^k E [    f(X^{(m)}_i) ] \leq \sum_{i=1}^k E [ f(X^{(m + 1)}_i) ],   
\end{equation*}\vspace{-0.1cm}
for all $0 \leq m \leq n - 1$. Hence,\vspace{-0.1cm}
\begin{equation} \label{theoremlast}
\sum_{i=1}^k E [ f(X^{(0)}_i) ] \leq \sum_{i=1}^k E [ f(X^{(n)}_i) ].
\end{equation}\vspace{-0.1cm}
Note that $(X^{(0)}_1, ..., X^{(0)}_k)$ are the final numbers of objects in bins $1, ..., k$ when all $n$ balls are distributed following RJ-CH, while $(X^{(n)}_1, ..., X^{(n)}_k)$ are the final numbers of objects in bins $1, ..., k$ when all $n$ balls are distributed following CH-BL.
Therefore \eqref{theoremlast} implies \eqref{theorem1.1}.
\hfill $\square$
\vspace{-0.1cm}
\subsection{Fewer bin searches}
\vspace{-0.1cm}
Bin searches are defined as the total number of bins (or servers) that must be searched to assign an object. It should be noted that this is not the total number of indexes in the array searched. We make this distinction because the latter tends to be implementation-specific. We will later provide experimental results for both metrics, but here we analyze object insertion as object removals in practice are taken care of by time-based decay and more stringent measures (Section \ref{appendix:objectRemoval}). Let the number of bin searches be denoted as $S$. Recall that there are $n$ objects, $k$ bins and a maximum capacity $C = \ceiling{(1 + \epsilon)\frac{n}{k}}$ for some $\epsilon \geq 0$.

\textbf{When inserting another object}, CH-BL achieves the following upper bounds on the expected value of $S$ as a function of $\epsilon$:\vspace{-0.1cm}
\begin{equation} \label{eq:BDsearches}
    f(\epsilon) = \begin{cases}
2 / \epsilon^2 & \text{$\epsilon < 1$ ,}\\
1 + \frac{log(1 + \epsilon)}{1 + \epsilon} & \text{$\epsilon \geq 1$ .}\\
\end{cases}
\end{equation}
\vspace{-0.1cm}
For RJ-CH, we assume a worst case scenario of $\floor{{n}/{C}}$ bins full, and we prove the following theorem.
\begin{theorem}
\label{lowerBinSearches}
Under RJ-CH, the expected value of $S$ is upper bounded by $1 + {1}/{\epsilon}$.
\end{theorem}\vspace{-0.1cm}

Observe that,\vspace{-0.1cm}
\begin{equation}
    f(\epsilon) = \begin{cases}
1 + \frac{1}{\epsilon} \ll 2 / \epsilon^2 & \text{$\epsilon < 1 $ and $\epsilon$ small ,}\\
1 + \frac{1}{\epsilon} \ll 1 + \frac{log(1 + \epsilon)}{1 + \epsilon} & \text{$\epsilon \geq 1$ and $\epsilon$ large .}\\
\end{cases}
\end{equation}

Setting a maximum capacity has a much greater impact for small $\epsilon$ and for small $\epsilon$, RJ-CH is an order of a magnitude better. For large $\epsilon$, RJ-CH is $log(1 + \epsilon)$ better. For $\epsilon$ slightly larger than 1, the methods are comparable. In practice, RJ-CH results in significantly fewer percentage of full bins which, in addition to the improved upper bound, results in an even more pronounced improvement in $S$.
\vspace{-0.1cm}
\subsection{Expected number of objects until first overflow}
\label{objectsTillFirstOverflow} \vspace{-0.1cm}
\emph{Stateless addressing} is one of the key requirements~\cite{SPOCA}, which is that the assignment process should be independent of the number of objects in the non-full bins. Methods that, for example, always assign new objects to the bin with the least objects are not viable for consistent hashing because keeping track of object distribution in a dynamic environment is too slow and requires costly synchronization.

In this section, we look at the expected number of objects that can be assigned before any bins are full. If all bins have the same capacity, then lower expected number of objects indicates poor load balancing since one of the servers was overloaded prematurely. RJ-CH produces the uniform distribution which is optimal under \emph{stateless addressing}~\cite{SPOCA}. 
Let $N_1$ be the number of objects assigned before any bin is full. \vspace{-0.1cm}
\begin{theorem}
\label{lowerFirstFullBin}
Both the probability of no full bin and $E[N_1]$ are maximized by the uniform distribution for all stateless addressing, which is achieved by RJ-CH.
\end{theorem} \vspace{-0.2cm}
\subsection{Lower initial bin load variance}
\label{varianceBeforeFirstOverflow}
\vspace{-0.1cm}
In this section we argue that even without the cascading effect, RJ-CH is still superior to the state-of-the-art. Recall that bin load is defined as the number of objects in a bin. Theorem \ref{lowerVarianceFirstFullBin} shows that RJ-CH minimizes bin load variance before the first full bin. This result applies over all distributions which satisfies the requirements of stateless addressing. Let $X_i$ be the random number of objects in bins $b_i$ and $p_i$ be the probability of an object being assigned to bin $b_i$. \vspace{-0.1cm}
\begin{theorem}
\label{lowerVarianceFirstFullBin}
Assume a fixed number of objects are assigned and no bins are full. $Var(X_i)$ is minimized by the uniform distribution for all stateless addressing, which is achieved by RJ-CH. 
\end{theorem} \vspace{-0.1cm}


Cascaded overflow starts when we hit the first full bin. Theorem \ref{lowerVarianceFirstFullBin} suggests that even before the start of the cascading effect, CH has poor variance compared to RJ-CH. This is important as even heavily loaded servers are undesirable practically.

\subsection{Object Assignment Probability Variance} \vspace{-0.1cm}
We define the object assignment probability of a bin as the probability that a new object lands in that bin. Note that this probability is dependent on the previous object assignments seen so far, and hence is a random variable. We are concerned with the variance of the object assignment probability for the non-full bins. We will use $p^j_i$ to refer to the random probability that a new object lands in the $i^{\text{th}}$ non-full bin when there are $j$ full bins. It should be noted that when there are $j$ full bins and $k$ total bins, we have $p^j_1,...,p^j_{k-j}$ assignment probabilities. The variance of this random variable, or the object assignment probability variance, is a measure of load balancing performance. In the ideal case with perfect load balancing, all assignment probabilities should be the same and the variance should be zero. It follows from universal hashing that RJ-CH has this property, with $p^j_1=...=p^j_{k-j}=1/(k-j)$. 
Therefore, we claim that RJ-CH is optimal in terms of this load balancing metric. CH-BL, on the other hand, has higher variance as it reassigns objects to the closest non-full bin in the clockwise direction. We obtain the following theorem: 

\begin{theorem}
\label{lowerObjectAllocationPVar}
Assume that each non-full bin has an equal probability of being full. For CH-BL, $Var(p^j_i)$ strictly increases exponentially with rate at least $1/(3k)$ for $j=1,...,k-3$.
\end{theorem}

The above theorem shows that the method of reassigning objects to the closest non-full bin in the clockwise direction is not only sub-optimal but also progressively worsens as more bins become full due to the cascading effect. We provide empirical results to support Theorem~\ref{lowerObjectAllocationPVar} in Appendix \ref{objectAssignmentProbabilityVariance}. \vspace{-0.1cm}

\section{Experimental Evaluations} \vspace{-0.2cm}
For evaluation, we provide both simulation results and results on real server logs. \vspace{-0.2cm}
\subsection{Simulation results}
We generate $n$ objects and $k$ bins where each bin has capacity $C = \ceiling{\frac{n}{k}(1 + \epsilon)} $. We hash each of the bins into a large array, resolving bin collisions by rehashing. Bins are populated according to the two methods of RJ-CH and CH-BL. We sweep $\epsilon$ finely between 0.1 and 3, performing 1000 trials from scratch for each $\epsilon$. We present results on percentage of bins full and wall clock time with 10000 objects and 1000 bins. Other results on variance of bin loads, bin searches, and objects till first full bin are given in Appendix \ref{appendixSimulationResults}. Another setting with less load is given in Appendix \ref{SupplementaryFigures}, and results are similar. Exact implementation details are given in Appendix \ref{implementationDetails}.

\begin{figure}[h!]
    \begin{minipage}[c]{0.4\linewidth}
    \includegraphics[width=\linewidth]{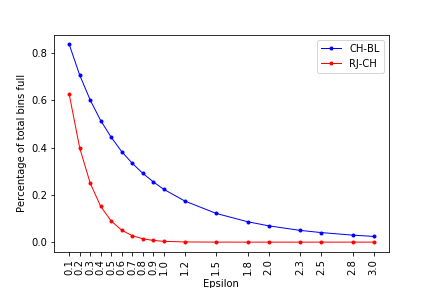}\vspace{-0.1cm}
  \caption{Percentage of total bins full. \newline \:}
    \label{fig:binsfull}
    \end{minipage}
    \hfill
    \begin{minipage}[c]{0.4\linewidth}
    
    \includegraphics[width=\linewidth]{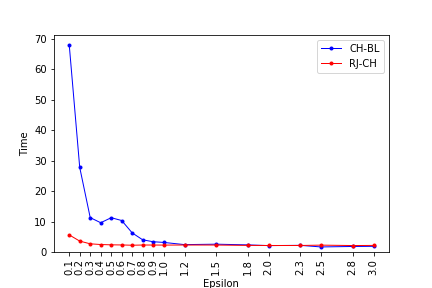}\vspace{-0.1cm}
    \caption{Wall clock time for adding $n + 1$th object.}
      \label{fig:time}
    \end{minipage}
\end{figure}

Figure \ref{fig:binsfull} shows the percentage of bins that are full. For most $\epsilon$, RJ-CH has a 20\% - 40\% lower percentage of \textit{total bins} that are full. For the case of $\epsilon = 0.3$, only 25\% of bins are full for RJ-CH as opposed to 60\% for CH-BL. Clearly, this implies that CH-BL causes servers to overload earlier than required, indicating poor load balancing.

Empirical results on the wall clock time for inserting the $n + 1$th object are given in Figure \ref{fig:time}. For wall clock time, RJ-CH attains between a 2x and 7x speedup for small $\epsilon$. The speedup results from the fewer number of full bins, practical considerations of hashing, and the cascaded overflow of CH-BL.
\vspace{-0.1cm}
\subsection{AOL search logs experiments} 
\vspace{-0.1cm}
In this section we present results with real AOL search logs. This is a dataset of user activity with 3,826,181 urls clicked, of which 607,782 are unique. We selected a wide range of configurations, Appendix \ref{appendixConfigurations} (Table \ref{aolConfig}), used in practice, such as reflecting the 80\% of internet usage being video \cite{ciscoStudy}. 

\textbf{Definitions:}
\begin{itemize}[leftmargin=*] \vspace{-0.1in}
  \item \textbf{Cache size:} Following the definitions in \cite{vaha2016bounded, CHBDblog} and implemented in practice for Vimeo \cite{vimeo} and Google \cite{CHBDblog}, cache size is defined as the maximum number of objects a cache server can hold. \vspace{-0.1cm}
  \item \textbf{Time-based eviction:} Stale urls are evicted after they have not been requested for a certain period of time. This is the most common eviction strategy in practice \cite{cachingASP, cachingMicrosoft, cachingMozilla}. \vspace{-0.1cm} 
  \item \textbf{Cache miss:} A cache miss is defined as a request for an object from a non-full bin where it has not already been cached \cite{cachingMicrosoft,karkhanis2002a}. This captures the resource intensive process of the cache server requesting and caching the object from a main server.\vspace{-0.1cm}
\end{itemize}
\vspace{-0.1cm}
Results are evaluated in cache misses, given in Table \ref{aolResults}. Cache misses are presented as additional cache misses, since there is a large number of unavoidable cache misses for a given eviction time even with no server failures and infinite capacity. In all configurations, RJ-CH significantly decreases the number of cache misses by several orders of magnitude.

\begin{table} \vspace{-0.2in}
\parbox{.45\linewidth}{
\centering
\caption{Additional cache misses on AOL search dataset.}
\begin{tabular}{ c|c|c } 
 \toprule
 Configuration & CH-BL & RJ-CH \\ 
 \midrule
 Config 1 & 35780 & \textbf{312} \\ 
 Config 2 & 52403 & \textbf{4680} \\ 
 Config 3 & 12223 & \textbf{104} \\ 
 Config 4 & 48571 & \textbf{9} \\ 
 \bottomrule
\end{tabular}
\label{aolResults}
}
\hfill
\parbox{.45\linewidth}{
\centering
\caption{Additional cache misses on Indiana University Clicks dataset.}
\begin{tabular}{ c|c|c } 
 \toprule
 Configuration & CH-BL & RJ-CH \\ 
 \midrule
 Config 1 & 72989 & \textbf{5549} \\ 
 Config 2 & 98712 & \textbf{9054} \\ 
 Config 3 & 105499 & \textbf{8641} \\ 
 Config 4 & 49304 & \textbf{3498} \\ 
 \bottomrule
\end{tabular}
\label{clicksResults}
}\vspace{-0.17in}
\end{table}


\vspace{-0.1cm} 
\subsection{Indiana University Clicks search logs experiments}
\vspace{-0.1cm} 
In this section we present results using Indiana University Clicks search logs. This is a dataset of user activity, where we use the first 1,000,000 urls clicked of which 26,062 are unique. For this dataset, we again selected a wide range of configurations used in practice, given in Appendix \ref{appendixConfigurations} (Table \ref{clicksConfig}). 

Again, results are evaluated in cache misses, given in Table \ref{clicksResults}. In all configurations, RJ-CH significantly decreases the number of cache misses by roughly one order of magnitude. \vspace{-0.2cm}

\section{Conclusion}\vspace{-0.1cm}
From both theoretical and empirical results, RJ-CH significantly improves on the state-of-the-art for dynamic load balancing. With this method, objects are much more evenly distributed across bins and bins rarely hit maximum capacity. In terms of bin load, we also prove the stochastic dominance of CH-BL over RJ-CH and a corollary is RJ-CH has lower expected number of full bins and bin load variance. On the AOL search dataset and Indiana University Clicks dataset with real user data, RJ-CH reduces cache misses by several orders of magnitude.

\section{Broader Impacts}
CH is widely used in industry including the popular chat app Discord with over 250 million users \cite{discord}, Amazon's storage system Dynamo \cite{Dynamo}, the distributed database system Apache Cassandra \cite{Cassandra}, Google's cloud system \cite{CHBDblog}, Vimeo's video streaming service \cite{vimeo}, and many others. Improved CH has a direct and significant impact reducing energy consumption, improving the latency of the services, and reducing server costs due to the popularity and widespread use of the aforementioned services.


\bibliography{references}

\newpage

\begin{appendices}
\section{Proof of Theorem \ref{mainTheorem}} \label{corollaryAppendix}
{\bf Theorem \ref{mainTheorem} restated } \ {\it  Following the notations in Theorem \ref{theorem1}, for $d \geq 1$, 
\begin{equation} \label{corollary1}
 E[  ( X^{(RJ-CH)}_i )^d ] \leq  E [  (  X^{(CH-BL)}_i )^d ], \qquad \hbox{for $i=1,..., k$}.
\end{equation}
In particular, 
\begin{equation}  \label{corollary2}
  E[ ( X^{(RJ-CH)}_i )^2 ]\leq   E[ (  X^{(CH-BL)}_i )^2 ] \text{ and }
var( X^{(RJ-CH)}_i )  \leq   var( X^{(CH-BL)}_i ) ,
\hbox{for $i=1,..., k$}.
\end{equation}
Moreover, 
\begin{equation} \label{corollary3}
E(L^{(RJ-CH)}) \leq E(L^{(CH-BL)}), 
\end{equation}
where  $L^{(CH-BL)}$ ($L^{(RJ-CH)}$)  is the number of full bins following the CH-BL (RJ-CH) method.
}\vspace{-0.1cm}

{\bf Proof of Theorem \ref{mainTheorem}} \ In \eqref{theorem1.2}, choose  $f(x) = x^d$ with $d \geq 1$, which is a convex function
on $\{0, 1, ..., C\}$.
Then \eqref{corollary1} follows. Observe that 
$ E(X_i^{(D\text{-}CH)}) = E(X_i^{(CH-BL)}) = n/k$. Then \eqref{corollary2} holds. 
Set $f(x) = I(x=C) $, which is also a convex function on $\{0, 1, ..., C\}$. Then \eqref{theorem1.1} 
implies \eqref{corollary3}.  
The proof is complete. \hfill  $\square$

\section{Proof of Lemma \ref{lemma1}}
It suffices to show that, for two schemes indexed by $m$ and $m + 1$, the final joint distributions of the numbers of objects in the $K$ bins are the same. Let $x$ be the number of objects in bin 1 before the assignment of the $m$th object. There are three cases.
  
Case 1. $x \leq C-2$. The two schemes give same distribution of the $m$th object and the $m+1$th object.
One will be assigned to bin 1 and the other uniformly distributed over the non-full bins before the $m + 1$th object has been assigned.
  
Case 2. $ x= C-1$. The two schemes give the same distribution of the $m$th and $m+1$th object. One object is added to bin 1 making it full and the other object is uniformly distributed over the rest of the non-full bins.
  
Case 3. $x = C$.  As bin 1 is full before the $m$th object has been assigned, both schemes will distribute the $m$th and $m+1$th objects uniformly to the non-full bins, one after the other.
  
In summary, the two schemes give same joint distribution of the numbers of objects in $K$ bins after the $m+1$th object has been assigned. Starting from $m+2$th object, the two schemes are the same. As a result, the final joint distributions of the numbers of objects in $K$ bins will be the same for all schemes regardless of the value of the index $m$. 
The proof is complete. \hfill $\square$

\section{Proof of Lemma \ref{lemma2}}
The distributions can be understood as the result of dropping $n$ objects into $k$ bins randomly. We omit the details. \hfill $\square$

\section{Proof of Lemma \ref{lemma3}}
For $x \geq n/2$, we set $s = n - x  \leq n/2$. Let $r_i = p_i/q_i \leq 1$. Denote by ${\cal B}_j$ the collection  of all subsets of $\{1, ..., n\}$ with size $j$. The cardinality of ${\cal B}_j$ is ${ n \choose j }$. Write
\begin{eqnarray*}
P(\xi_1 = s) &=& \sum_{ \delta_1 + ... + \delta_n = s \atop \delta_i = 0, 1} 
p_1^{\delta_1} {\cdots} p_n^{\delta_n} q_1^{1-\delta_1} {\cdots} q_n^{1-\delta_n}
\\
&=& q_1 \cdots q_n \sum_{ \delta_1+ ... + \delta_n= s \atop \delta_i = 0, 1}
r_1^{\delta_1} {\cdots} r_n^{\delta_n}
\\
&=& q_1 \cdots q_n \sum_{ (i_1, ..., i_s) \in {\cal B}_s}  
r_{i_1} {\cdots} r_{i_s}
\\
&\geq & 
  q_1 \cdots q_n \sum_{ (i_1, ..., i_s) \in  {\cal B}_s}  
{ 1 \over {n-s \choose n-2s} }   \sum_{ (j_1, ..., j_{n-s} ) \in  {\cal B}_{n-s} \atop
(i_1, ..., i_{s} ) \subseteq (j_1, ..., j_{n-s} )  	}  
r_{j_1} {\cdots} r_{j_{n-s} }
\\
&=& 
q_1 \cdots q_n    \sum_{ (j_1, ..., j_{n-s} ) \in  {\cal B}_{n-s}  	}  
r_{j_1} {\cdots} r_{j_{n-s} } 
\sum_{(i_1, ..., i_s) \in  {\cal B}_s \atop 
	(i_1, ..., i_{s} ) \subseteq (j_1, ..., j_{n-s} )  	} { 1 \over {n-s \choose n-2s}}
\\
&=& 
q_1 \cdots q_n    \sum_{ (j_1, ..., j_{n-s} ) \in  {\cal B}_{n-s}  	}  
r_{j_1} {\cdots} r_{j_{n-s} } \times 1
\\
&=& P(\xi_1= n-s)
\\
&=& P(\xi_2= s).
\end{eqnarray*}
And $P(\xi_1 = x) =P(\xi_1 = n-s) \leq P(\xi_1 = s) = P(\xi_2 = x)$ for $x \geq n/2$. 
Then, \eqref{lemma3.1} holds and $\xi_1 \prec \xi_2$.
Since, for $x$ satisfying $x < C $ and $n - x < C$, 
\begin{equation*}
{P(\xi_1=x | \xi_1 < C, \xi_2 < C) \over P(\xi_2=x |\xi_1 < C, \xi_2 < C)} 
= {  P(\xi_1=x) \over P(\xi_2=x) },
\end{equation*}
\eqref{lemma3.2} follows. As a result, \eqref{lemma3.3} holds.  \hfill $\square$

\section{Proof of Lemma \ref{lemma4}}
For $K < k$, consider $D$ objects been uniformly assigned into a cluster of bins $1, ..., k$, and reassigned according to the CH-BL. Denote as $Q(D, K)$ the probability that no objects are reassigned to beyond bin $1$. In other words, $Q(D, K)$ is the probability that all $D$ objects are "self-contained" in bins $1, ..., K$ under the CH-BL of relocation.  
 
Fix values $z_1, ..., z_J$ such that $0 \leq z_j < C$ and $\sum_{j=1}^J z_j = n^*$. We consider the conditional distribution under the condition $L_{i_j} = l_j$ for some fixed $l_j$ , $j=1,..., J$.  Observe that $n^* = n- ( k - J) C = n - C\sum_{j=1}^J l_j$.
Write
\begin{eqnarray*}
 &&  P(x_{i_1}
  = z_1, ..., x_{i_J} = z_J | L_{i_1} = l_1, ..., L_{i_J}= l_J)
  \\
  &\propto & P(\hbox{ $Cl_j+z_j$ objects are assigned to $l_j+1$ bins, $j=1,... J$} ) \times
  \\&& 
 \prod_{j=1}^J \Bigl\{ P( \hbox{ $Cl_j$ objects are assigned to $l_j$ bins $\,\,|\,\,$ $Cl_j+z_j$ objects assigned to $l_j+1$ bins} ) \\
 &&
  \times P(\hbox{ $Cl_j$ objects assigned to the $l_j$ bins are "self-contained" under CH-BL} )
\Bigr\} 
 \\
 &\propto& { n \choose z_1 + l_1C, ..., z_J + l_J C} \bigl( { 1 + l_1 \over k} \bigr)^{ Cl_1 +z_1} \cdots 
  \bigl( { 1 + l_J \over k} \bigr)^{ Cl_J+z_J} \times
  \\
  && \prod_{j=1}^J \bigl\{ { z_j + Cl_j \choose z_j,  Cl_j} 
  \bigl( { l_j \over l_j +1 } \bigr)^{ Cl_j } Q( C l_j, l_j)
  \bigr\} 
  \\
  &=& { n \choose n^*, n-n^*} { n^* \choose z_1, ..., z_J} \bigl( { 1 + l_1 \over k} \bigr)^{ z_1 }\cdots 
  \bigl( { 1 + l_J \over k} \bigr)^{  z_J} \times 
  \\ && {n - n^* \choose Cl_1, ..., Cl_J} 
  \bigl( {  l_1 \over k} \bigr)^{ Cl_1  } \cdots 
  \bigl( {  l_J \over k} \bigr)^{ Cl_J } Q(Cl_1, l_1) \cdots Q(C l_J, l_J)
  \\
  & \propto &
  { n^* \choose z_1, ..., z_J} (  p_1^*    )^{ z_1 } \cdots ( p_J^*    )^{ z_J }.
\end{eqnarray*}
The proof is complete. \hfill $\square$

\section{Proof of Lemma \ref{lemma5}}
There are two key observations. First, 
\eqref{lemma5.1} is equivalent to minimization of $\sum_{j=1}^k E [ f(X_j) I( j\in {\cal N} )]$, since the full bins in Step 1 will remain full till the end. 
Second, if we change Step 3 to "placing $m$ balls into bins in $\cal N$ following RJ-CH", the distribution 
of $(X_1, ..., X_k)$ will not change. We next argue that
the distribution of $(X_1, ..., X_k)$ will not change, if we change the entire distribution scheme in Steps 1-3 to Steps (a)-(c) in the following:
\begin{enumerate}
	\item same as Step 1.
	\item same as Step 3.
	\item same as Step 2, and reassigning following RJ-CH. 
\end{enumerate} 
Steps (b) and (c) switch Steps 2 and 3. Unlike in Step 2, where the object need not be reassigned, in Step (c), the object may be assigned to a full bin and, in that case, reassigned following RJ-CH.

The equivalence of these two schemes of object assignment, one described in Steps 1-3 and one in Steps (a)-(c), in terms of the distribution of $(X_1,..., X_k)$, can be understood by tracking the object in Step 2 and that in Step (c). Suppose in Step 2, the object is assigned to some bin $i_j$. It follows from Lemma 3 that the final distribution of the numbers of objects in the $k$ bins will not change if the object in Step 2 is instead assigned as the last object into bin $i_j$ and reassigned following RJ-CH. Since both happen with same probability $q_j$, the desired equivalence holds true.  As a result, it suffices to prove \eqref{lemma5.1} for the object distribution scheme in Steps (a)-(c).

Recall that ${\cal N} = \{i_1,...,i_J\}$ are the non-full bins after Step (a). Let $\tilde {\cal N} = \{ \tilde i_1, ..., \tilde i_s\} \subseteq {\cal N} $ be all the non-full bins  after Step (b) with $\eta_j$ denoting the number of objects in bin $\tilde i_j$. Clearly $s \leq J$. We show that, conditioning on $\tilde {\cal N}  $ and $L_{i_j}, j \in {\cal N}$,  
\begin{equation} \label{lemma5.2}
\eta_1 \prec \cdots \prec \eta_s
\end{equation}
where $\prec$ means stochastically smaller. For ease of notation and without loss of generality, let $\tilde i_1 = i_1, ..., \tilde i_s = i_s$. 

For simplicity of exposition, we only show the conditional stochastic dominance: $\eta_1 \prec \eta_2$. 
Let $a_1$ and $a_2 $ be two nonnegative integers. Let $\Delta_j $, $j=1, ..., a_1+a_2$,  be independent random variables taking values $1$ and $0$, with probabilities such that $P(\Delta_j =1 ) = p_1^*/(p_1^* + p_2^*) = 1-P(\Delta_j=0)$ for $j=1,..., a_1$ and $P(\Delta_j = 1) =  P(\Delta_j=0) = 1/2 $ for $j=a_1+1, ..., a_1+a_2$, where $p_l^* = (1+L_{i_l})/K$. Set $\xi_1 = \sum_{j=1}^{a_1+a_2} \Delta_j$ and $\xi_2 = a_1+ a_2 - \xi_1$. Since $p_1^* \leq p_2^*$, Lemma 2 implies that,  
\begin{equation*}
\xi_1 \, |  \, (\xi_1 < C, \xi_2 <C ) \prec \xi_2 \,|  \, (\xi_1 < C, \xi_2 <C).
\end{equation*} 
Now consider the condition that, in Step (a), there are a total of $a_1$ objects in bins $\tilde i_1$ and $\tilde i_2$ and in Step (b), there are a total of $a_2$ additional objects in bins $\tilde i_1$ and $\tilde i_2$. It follows from Lemmas 1 and 4 that, under this condition, the conditional distribution of $(\eta_1, \eta_2)$ is the same as the conditional distribution of the above $(\xi_1, \xi_2)$, under the condition that $\xi_1 <C $ and $\xi_2 < C$. As a result, the conditional stochastic dominance of $\eta_1 \prec \eta_2$ in \eqref{lemma5.2} is proved. 

Recall that we set $\tilde i_1= i_1,..., \tilde i_s=i_s$ for notational convenience. After Step (c), the number of objects in bin $\tilde i_j$, $j=1,..., s$, is 
\begin{equation*}
\tilde \eta_j = \begin{cases}
 \eta_j +1  & \hbox{with conditional probability } q_{ j} +1/s - \bar q_s 
	\\
	\eta_j  & 
	\hbox{with conditional probability }  1- q_{ j} -  1/s  + \bar q_s 
	\end{cases}
\end{equation*} 

where $\bar q_s = \sum_{j=1}^s q_j/s$ and the conditioning is on $\tilde {\cal N} ,  L_{i_j}, j \in {\cal N}$. Here $q_j$ is the probability the last object is assigned to bin $i_j$ and $1/s - \bar q_s$ is the probability the object is assigned to the full bins $i_{s+1},..., i_{J}$, with probability $1-\sum_{j=1}^s q_j$, then reassigned to bin $i_j$. 

Observe that, since $f$ is convex, $f(x+1) -f(x)$ is an increasing function of $x \in \{  0, 1, ..., C-1\}$. Hence \eqref{lemma5.2} implies  $E[f(\eta_j+1)- f(\eta_j)  \, |\,  \tilde {\cal N} ,  L_{i_j}, j \in {\cal N}]$ is  increasing in $j$. Since $q_j$, $j=1,...J$, are   monotone increasing in $j$, it follows that
\begin{eqnarray*}
 && E\Bigl\{ \sum_{j=1}^s [  f(\tilde \eta_j) - f(\eta_j) ]  \, |\,  \tilde {\cal N} ,  L_{i_j}, j \in {\cal N} \Bigr\} 
 \\
 &=& E\Bigl\{ \sum_{j=1}^s [  f(  \eta_j +1) - f(\eta_j) ][ q_j + 1/s - \bar q_s ]  \, |\,  \tilde {\cal N} ,  L_{i_j}, j \in {\cal N} \Bigr\} 
 \\
 &=&  \sum_{j=1}^s [ q_j  - \bar q_s ] E  [  f( \eta_j +1) - f(\eta_j)  \, |\, \tilde {\cal N} ,  L_{i_j}, j \in {\cal N} ] 
 \\
 && 
 + (1/s) \sum_{j=1}^s E[ f( \eta_j +1) - f(\eta_j)  \, |\,  \tilde {\cal N} ,  L_{i_j}, j \in {\cal N} ]
 \\
 &\geq& 
  (1/s) \sum_{j=1}^s E[ f(  \eta_j +1) - f(\eta_j)   \, |\, \tilde {\cal N} ,  L_{i_j}, j \in {\cal N} ].
\end{eqnarray*}
where, in the last inequality, the equality holds when all $q_j$ are equal, i.e.,  $q_1=...=q_J= 1/J$. This inequality holds because the correlation of two sequences of increasing numbers is always nonnegative. Therefore, the conditional mean of $  \sum_{j=1}^s f(\tilde \eta_j)   $ is minimized  when $q_1=...=q_J= 1/J$. Since $(\tilde \eta_1, ..., \tilde \eta_s)$ are the final numbers of the objects in bins $\{\tilde i_1, ..., \tilde i_s\}$ and the rest of the bins are already full after Step (b), we conclude that $E [(\sum_{j=1}^k f(X_j)]$ is minimized when $q_j$ are all equal. \eqref{lemma5.1} is proved. \hfill $\square$

\section{Proof of Theorem \ref{lowerBinSearches}}
Recall that there are $n$ objects, $k$ bins, and capacity $C = (1 + \epsilon)\frac{n}{k}$ for some $\epsilon \geq 0$. Our claim is the RJ-CH method of assigning objects with uniform distribution to the non-full bins is expected to search $1 + {1}/{\epsilon}$ bins to assign an object to a non-full bin in the worst case scenario. To show the upper bound, we assume the worst case scenario of $\floor{{n}/{C}}$ bins full. Then,
\begin{equation}
    \frac{1}{1-{\floor{\frac{n}{C}}}/{k}} = \frac{k}{k - \floor{\frac{n}{C}}} \leq \frac{k}{k - \frac{n}{C}} = \frac{k}{k - \frac{n}{(1 + \epsilon)n / k}} = \frac{k}{k - \frac{k}{1 + \epsilon}} = 1 + \frac{1}{\epsilon}
\end{equation}
The proof is complete. \hfill $\square$

\section{Proof of Theorem \ref{lowerFirstFullBin}}
Recall that there are $n$ objects, $k$ bins, and capacity $C = (1 + \epsilon)\frac{n}{k}$ for some $\epsilon \geq 0$. $N_1$ is the number of objects assigned before any bin is full. Our claim is the RJ-CH method of assigning objects with uniform distribution to the non-full bins maximizes both the probability no bin is full and $E[N_1]$.

For the binomial case where $b_1$ refers to bin 1 and $b_2$ refers to bin 2, $Pr[b_1, b_2 \text{ not full }| m\text{ objects in $b_1, b_2$}, p_1, p_2]$ is uniquely maximized by $p_1 = p_2 = 1/2$, where $C \leq m \leq 2(C - 1)$. 

For the multinomial case, where $C \leq n \leq k(C - 1)$ and bins $b_1,...,b_k$ have probability $p_1,...,p_k$:

\begin{align*}
    & {Pr}[\text{No bin full | $n$ objects, $k$ bins}] 
    \\ = & \sum_m{Pr}[\text{$m$ objects in $b_i, b_j$}]
    \\ & \times {Pr}[b_i, b_j \text{ not full | $m$ objects in $b_i, b_j$}]
    \\ & \times {Pr}[\text{\{$b_1,...,b_k$\} - \{$b_i, b_j$\} not full | $n - m$ objects in \{$b_1,...,b_k$\} - \{$b_i, b_j$\}}]
\end{align*}

If we consider the term $Pr[b_i, b_j \text{ not full | $m$ objects in $b_i, b_j$}]$ then
\begin{align*}
    & Pr[b_i, b_j \text{ not full | $m$ objects in $b_i, b_j$, $p_i, p_j$}]
    \\ \leq & \: Pr[b_i, b_j \text{ not full | $m$ objects in $b_i, b_j$, $p_i = \frac{p_i + p_j}{2}, p_j = \frac{p_i + p_j}{2}$}]
\end{align*}

with strict inequality when $m \geq C$. Thus, for $n \geq C$ every pair of bin probabilities can be repeatedly replaced by their mean, and $Pr[\text{No bin full | $n$ objects $k$ bins}] $ is then uniquely maximized by $p_1 = ... = p_k = {1}/{k}$. 

Recall that $N_1$ is the number of objects assigned before any bin is full.

\begin{align*}
    E[N_1] & = \sum_{l=0}^{n} Pr[N_1 > l] = \sum_{l=0}^{n} Pr[\text{no bin full | $l$ objects $k$ bins}]
\end{align*}

Each term is maximized by the uniform distribution and the proof is complete. \hfill $\square$

\section{Proof of Theorem \ref{lowerVarianceFirstFullBin}}
To see that RJ-CH minimizes bin load variance before the first full bin, we only need to show that the conditional second moment of bin load is minimized by RJ-CH
since the conditional mean is fixed as $n/k$ for each bin.  Consider bins $b_i$ and $b_j$. Let $X_i$ ($X_j$) be the random number of objects in bins $b_i$ ($b_j$)  and $p_i$ ($p_j$) be the probability of objects being assigned to bin $b_i$ ($b_j$). Assume $X_i+X_j= z$ for any fixed number $z$. Under this condition, the conditional probability of an object assigned to $b_i$ given it is in $b_i$ or $b_j$, is $p= p_i/(p_i+p_j)$. 

Suppose $X$ and $Y$ are two random variables following binomial distributions with number of trials as $z$ and parameter as $p$ and $1/2$ respectively. 
Let $X^*=\max(X, z-X)$ and $Y^*= \max(Y, z-Y)$. 
Then $Pr(X^*=y) = {\binom{z}{y}}[p^y(1-p)^{z-y} + (1-p)^y p^{z-y}] $ for $z \geq y> z/2$, and, if $z$ is even, $Pr(X^*=z/2) = {\binom{z}{z/2}}[p (1-p)]^{z/2}$. $Pr(Y^*=y)$ has an identical expression with $p=1/2$.  
A key observation is that the ratio of the probability functions of $X^*$ and $Y^*$ is increasing. Namely, 
\[  Pr(X^*=y)/Pr(Y^*=y) = (p^y(1-p)^{z-y} + (1-p)^y p^{z-y})/(2 \times 0.5^z) \] 
as a function of $y$ in the region $[z/2, z]$ is convex and increasing. It then follows that $X^*$ is stochastically greater than $Y^*$, i.e., $P(X^* \geq C) \leq P(Y^* \geq C)$, or,
equivalently, $P(X^* < C) \geq P(Y^* < C)$ for any $C$.

Moreover, the monotone increasing ratio of probability functions of $X^*$ and $Y^*$ also implies, for any $C$, the conditional distribution of $X^*$ given $X^*<C$ and the conditional distribution of $Y^*$ given $Y^*<C$ still has a monotone increasing ratio of probability functions. Hence, the conditional distribution of $X^*$ given $X^*<C$ is still stochastically greater than the conditional distribution of $Y^*$ given $Y^*< C$. As a result, $E( g( X^*)  | X^* < C) \geq E(g(Y^*)  | Y^* < C)$ for any increasing function $g$. Choose  $g(x) = x^2 + (z-x)^2$, which is increasing in $x$ on $[z/2, z]$.
Then, 
\begin{align*}
  & E( X^2 +(z-X)^2   | X^* < C) = E( ( X^*)^2 +(z-X^*)^2   | X^* < C) 
  \\
  \geq & \: E( ( Y^*)^2 +(z-Y^*)^2   | Y^* < C) =E( Y^2 +(z-Y)^2   | X^* < C).
\end{align*}

The above argument implies, conditioning on bins $i$ and $j$ having a total of $z$ objects, the conditional mean of $X_i^2 + X_j^2$ is minimized by the uniform distribution.

Thus considering all $k$ bins, to minimize the bin load variance before the first full bin repeatedly replace every pair of bin probabilities by their mean. The proof is complete. \hfill $\square$

\section{Proof of Theorem \ref{lowerObjectAllocationPVar} and a simulation} \label{objectAssignmentProbabilityVariance}

Recall that $p^j_1,...,p^j_{k-j}$ are random variables that represent probabilities where $p^j_i$ is the probability an object lands in the $i$th non-full bin with $j$ full bins. Note that $p^j_1,...,p^j_{k-j}$ are identically distributed. There are $k$ total bins and for RJ-CH $p^j_1=...=p^j_{k-j}=1/(k-j)$. We define the object assignment probability variance as $Var(p^j_i)$. Clearly, for RJ-CH $Var(p^j_i) = 0$. The method of CH-BL reassigns objects that attempt to be assigned to a full bin to the closest non-full bin in the clockwise direction. 

Let $p^{j-1}_{full}$ be the random variable that refers to the object assignment probability of the $j-1$th bin to be full. Consider $p^j_i$,

\begin{equation*}
    p^j_i = \begin{cases}
p^{j-1}_i + p^{j-1}_{full} & \text{with probability $\frac{1}{k-j}$ ,}\\
p^{j-1}_i & \text{with probability $1 - \frac{1}{k-j}$ .}\\
\end{cases}
\end{equation*}

Observe that,
\begin{align*}
    E[p^j_i | p^{j-1}_1,...,p^{j-1}_{k - (j-1)}] = & \:
    \frac{1}{k-j} (p^{j-1}_i + p^{j-1}_{full}) + (1 - \frac{1}{k-j}) (p^{j-1}_i) = p^{j-1}_i + \frac{p^{j-1}_{full}}{k-j} \text{ ,}
\end{align*}
and
\begin{align*}
    Var(p^j_i | p^{j-1}_1,...,p^{j-1}_{k - (j-1)}) = &\: \frac{1}{k-j} ((p^{j-1}_i + p^{j-1}_{full}) - E[p^j_i | p^{j-1}_1,...,p^{j-1}_{k - (j-1)}] )^2 
    \\
    & + (1 - \frac{1}{k-j})(p^{j-1}_i - E[p^j_i | p^{j-1}_1,...,p^{j-1}_{k - (j-1)}])^2
    \\
    = & \: \frac{1}{k-j} (p^{j-1}_{full} - \frac{p^{j-1}_{full}}{k-j})^2 + (1 - \frac{1}{k-j})(\frac{p^{j-1}_{full}}{k-j})^2
    \\
    = & \: \frac{1}{k-j} (1 - \frac{1}{k-j})^2 (p^{j-1}_{full})^2 + (1 - \frac{1}{k-j}) (\frac{1}{k-j})^2 (p^{j-1}_{full})^2
    \\
    = & \: \frac{1}{k-j} (1 - \frac{1}{k-j}) (p^{j-1}_{full})^2 \text{ .}
\end{align*}
We can write

\begin{align*}
    Var(p^j_i) & = E[Var(p^j_i | p^{j-1}_1,...,p^{j-1}_{k - (j-1)})] + Var(E[p^j_i | p^{j-1}_1,...,p^{j-1}_{k - (j-1)}])
    \\
    & = E[\frac{1}{k-j} (1 - \frac{1}{k-j}) (p^{j-1}_{full})^2] + Var(p^{j-1}_i + \frac{p^{j-1}_{full}}{k-j})
    \\
    & = E + V \text{, say.}
\end{align*}

Since $p_1^{j-1} + ... + p_{k - (j - 1)}^{j - 1} = 1$ and $p_1^{j-1}, ... ,p_{k - (j - 1)}^{j - 1}$ follow the same distribution, it follows that

\begin{align*}
    0 & = Var(\sum_{i = 1}^{k - (j - 1)}p_i^{j-1})
    \\
    & = \sum_{i= 1}^{k - (j - 1)}(Var(p_i^{j-1})) + \sum_{i \neq \tilde{i}} Cov(p_i^{j-1}, p_{\tilde{i}}^{j-1})
    \\
    & = (k - (j - 1))(Var(p_i^{j-1})) + ((k - (j - 1))^2 - (k - (j - 1))) Cov(p_i^{j-1}, p_{\tilde{i}}^{j-1}) \text{ .}
\end{align*}

As a result, 
\begin{align*}
    Cov(p_i^{j-1}, p_{\tilde{i}}^{j-1}) & = -\frac{1}{k - (j - 1) - 1} Var(p_i^{j-1}) \text{ .}
\end{align*}

Therefore, we can see that the V term is given by
\begin{align*}
    Var(p^{j-1}_i + \frac{p^{j-1}_{full}}{k-j}) & = Var(p^{j-1}_i) + (\frac{1}{k - j})^2 Var(p^{j-1}_{full}) + \frac{2}{k - j} Cov(p^{j-1}_i, p^{j-1}_{full})
    \\
    & = (1 + \frac{1}{(k - j)^2} + \frac{2}{k-j}(-\frac{1}{k-(j - 1) - 1}))Var(p^{j-1}_i)
    \\
    & = (1 - \frac{1}{(k - j)^2}) Var(p^{j-1}_i) \text{ .}
\end{align*}

The E term is given by
\begin{align*}
    E & = E[\frac{1}{k-j} (1 - \frac{1}{k-j}) (p^{j-1}_{full})^2]
    \\
    & > (\frac{1}{k-j} - \frac{1}{(k-j)^2}) Var(p_i^{j-1}) \text{ .}
\end{align*}

Combining the two terms, we have 
\begin{align*}
    E + V & > (1 + \frac{1}{k - j} - \frac{2}{(k - j)^2}) Var(p_i^{j-1}) \text{ .}
\end{align*}

For $k - j \geq 3$, ${1}/{k - j} - {2}/{(k - j)^2} > 0$ and for $k - j = 2$, ${1}/{k - j} - {2}/{(k - j)^2} = 0$.
Therefore, $Var(p^j_i)$ strictly increases for $j = 1,...,k-2$ and increases geometrically with rate at least $1/(3k)$ for $j=1,...,k-3$.
The proof is complete. \hfill $\square$

\begin{figure}[h!]
  \centering
  \begin{subfigure}[b]{0.75\linewidth}
    \includegraphics[width=\linewidth]{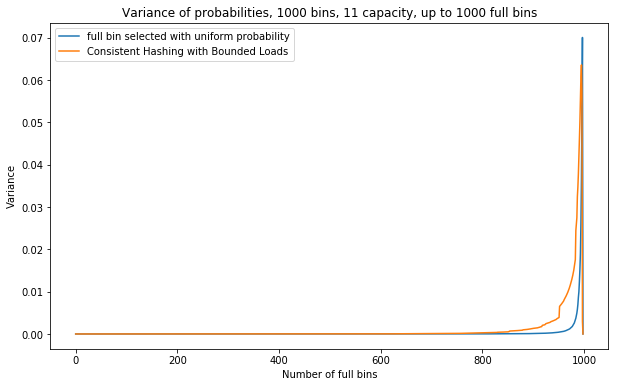}
    \caption{1000 bins, 11 capacity.}
  \end{subfigure}
  \begin{subfigure}[b]{0.75\linewidth}
    \includegraphics[width=\linewidth]{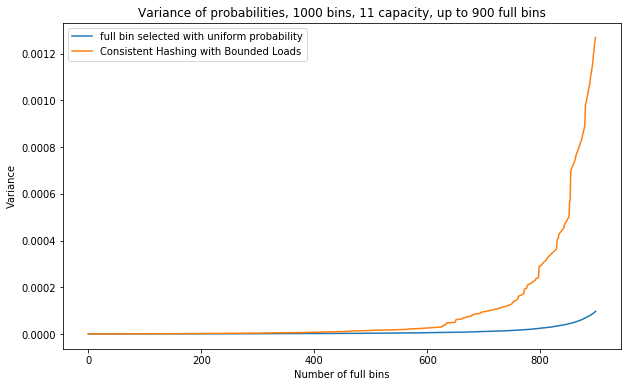}
    \caption{1000 bins, 11 capacity, up to the 900th full bin.}
  \end{subfigure}
    \begin{subfigure}[b]{0.75\linewidth}
    \includegraphics[width=\linewidth]{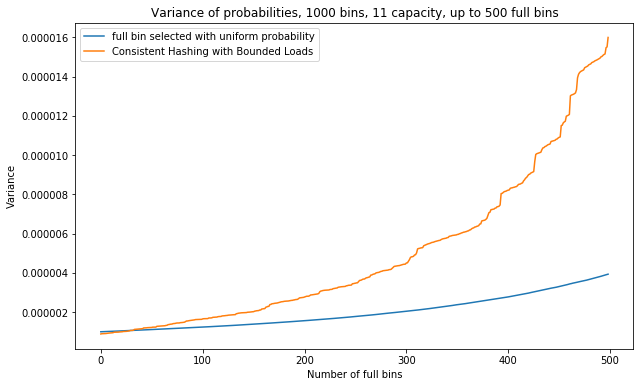}
    \caption{1000 bins, 11 capacity, up till the 500th full bin.}
  \end{subfigure}
  \caption{Variance of object assignment probabilities against number of full bins.}
  \label{fig:objectAllocProbVar}
\end{figure}

A simulation is given in Figure \ref{fig:objectAllocProbVar} where we also include CH-BL. The configuration for CH-BL is 1000 bins and capacity 11 ($\epsilon = 0.1$). We run 100 trials and take the mean. CH-BL performs significantly worse without the assumption that each non-full bin has equal probability of being the next full bin.

\newpage
\section{AOL search logs and Indiana University Clicks configurations} \label{appendixConfigurations}
Configurations for the AOL search dataset are given in Table \ref{aolConfig}. Configurations for the Indiana Clicks dataset are given in Table \ref{clicksConfig}.
\begin{table}
\vspace{-0.25in}
\caption{Distributed caching configuration for AOL search dataset.}
\begin{center}
\begin{tabular}{ c|c|c|c|c } 
 \toprule
 Setting & Config 1 & Config 2 & Config 3 & Config 4 \\ 
 \midrule
 Number of servers & 150 & 1000 & 100 & 20\\ 
 Cache size & 100 & 15 & 100 & 300 \\
 Minutes for stale urls to be evicted & 300 & 300 & 120 & 120 \\
 Minutes requests are served & 10 & 10 & 5 & 3\\
 Minutes for failed server to recover & 20 & 10 & 10 & 10\\
 Number of concurrent requests till server failure & 50 & 15 & 50 & 500\\
 \bottomrule
\end{tabular}
\end{center}
\label{aolConfig}
\end{table}

\begin{table}
\vspace{-0.2in}
\caption{Distributed caching configuration for Indiana University Clicks dataset.}
\begin{center}
\begin{tabular}{ c|c|c|c|c } 
 \toprule
 Setting & Config 1 & Config 2 & Config 3 & Config 4 \\ 
 \midrule
 Number of servers &  500 & 1000 & 800 & 200 \\ 
 Cache size & 500 & 300 & 300 & 3000 \\
 Minutes for stale urls to be evicted & 30 & 120 & 15 & 30 \\
 Minutes requests are served & 5 & 5 & 5 & 3 \\
 Minutes for failed server to recover & 10 & 10 & 7 & 15 \\
 Number of concurrent requests till server failure & 2000 & 1000 & 1000 & 5000 \\
 \bottomrule
\end{tabular}
\end{center}
\label{clicksConfig}
\end{table}
\newpage
\section{Additional simulation results} \label{appendixSimulationResults}
We generate $n$ objects and $k$ bins where each bin has capacity $C = \ceiling{\frac{n}{k}(1 + \epsilon)} $. We hash each of the bins into a large array, resolving bin collisions by rehashing. Bins are populated according to the two methods of RJ-CH and CH-BL.

We present results here with 10000 objects and 1000 bins, and also show results with less load, 3000 objects and 1000 bins, in Appendix~\ref{SupplementaryFigures}. Results in this different setting are similar.
We draw attention to the 10000 objects, 1000 bins case as low object bin ratios tend to be easier in practice. However, even in a 1:1 object to bin ratio, we observe that RJ-CH achieves superior load balance. The larger the object to bin ratio, the better RJ-CH is in comparison. 

We also performed the same simulations allowing objects and bins to arrive and leave. After placing all $n$ objects, objects and bins arrive and leave at a rate of $n/k$ of objects to bins. We then observe the load balancing metrics after $n$ objects have arrived or left. The results with such a methodology are similar to the previously introduced simulation methodology and we omit these results for succinctness. For additional experiments with a variety of combinations of configurations, see Appendix \ref{SupplementaryFigures}. For implementation details, see Appendix \ref{implementationDetails}. 

\subsection{Variance of bin loads}

\begin{table}
\vspace{-0.4in}
\caption{Mean and standard deviation of variance of bin loads and percentage of bins full}
\begin{center}
\begin{tabular}{ |c|c|c|c|c|c|c|c|c| } 
 \toprule
 & \multicolumn{4}{c}{Variance of bin loads} & \multicolumn{4}{c}{Percentage of bins full} \\ \midrule
 $\epsilon$ & CH-BL mean & std & RJ-CH mean & std & CH-BL mean & std & RJ-CH mean & std \\ 
 \midrule
 0.1 & 6.8 & 0.2 & \textbf{2.6} & 0.1 & 0.837 & 0.006 & \textbf{0.626} & 0.010\\ 
 0.3 & 19.1 & 0.4 & \textbf{6.6} & 0.2 & 0.602 & 0.009 & \textbf{0.250} & 0.010 \\ 
 1 & 51.9 & 1.2 & \textbf{10.0} & 0.4 & 0.224 & 0.009 & \textbf{0.003} & 0.002\\
 3 & 95.0 & 3.6 & \textbf{10.0} & 0.5 & 0.024 & 0.004 & \textbf{0.000} & 0.000\\ 
 \bottomrule
\end{tabular}
\end{center}
\label{variancetablemain}
\end{table}

Recall that bin load is defined as the number of objects in a bin. Figure \ref{fig:variance} shows the variance of bin loads against $\epsilon$ for different object bin configurations. A small variance is an indicator of better load balancing. RJ-CH achieves a 3x-10x improvement in bin load variance for small and large $\epsilon$, with tabulated data in Table \ref{variancetablemain}.


\subsection{Bin searches}

\begin{figure}[h!]

    \begin{minipage}[c]{0.45\linewidth}
    
    \includegraphics[width=\linewidth]{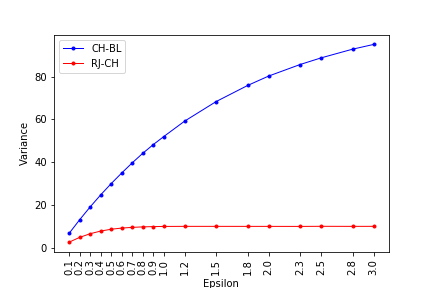}
      \caption{Variance of bin loads. \newline \:}
      \label{fig:variance}
  
    \end{minipage}
    \hfill
    \begin{minipage}[c]{0.45\linewidth}
    \includegraphics[width=\linewidth]{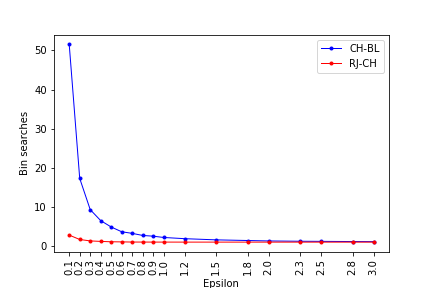}
  \caption{Bin searches to assign the $n + 1$th object.}
  \label{fig:searches}
    \end{minipage}
\end{figure}

Recall that bin search is defined as the number of bins searched. Figure \ref{fig:searches} shows the bin searches for the $n + 1$th object to be assigned after $n$ objects and $k$ bins have already been placed. Large $\epsilon$ is uninteresting for this case as there will be very few full bins. For interesting $\epsilon$, RJ-CH achieves a staggering 10x-25x improvement in bin searches. 

\vspace{-0.1in}

\vspace{-0.1in}

\subsection{Percentage of bins full}


Figure \ref{fig:binsfull} shows the percentage of bins that are full. For most $\epsilon$, RJ-CH has a 20\% - 40\% lower percentage of \textit{total bins} that are full. For the case of $\epsilon = 0.3$, only 25\% of bins are full for RJ-CH as opposed to 60\% for CH-BL, given in Table \ref{binsfulltablemain}. Clearly, this implies that CH-BL causes servers to overload earlier than required, indicating poor load balancing.

\subsection{Objects till first full bin}
\begin{figure}[h!]
    \begin{minipage}[c]{0.45\linewidth}

    \includegraphics[width=\linewidth]{mainManuscriptGraphs/PctBinsFull10000.png}
  \caption{Percentage of total bins full. \newline \:}
    \label{fig:binsfull2}
    \end{minipage}
    \hfill
    \begin{minipage}[c]{0.45\linewidth}
          \includegraphics[width=\linewidth]{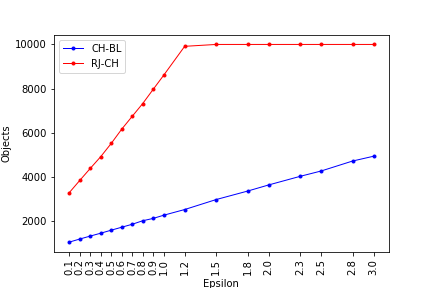}
  \caption{Number of objects assigned until one bin is full.}
  \label{fig:objectsTillFirstFull}
    \end{minipage}
\end{figure}

Figure \ref{fig:objectsTillFirstFull} shows the number of objects that are assigned before a bin is full. This number indicates the amount of load the system can tolerate before observing an overloaded server. RJ-CH achieves a 3x-5x improvement in the number of objects until one bin is full for both small and large $\epsilon$. 

\subsection{Steps and wall clock time}
\label{stepsAndWallClockTime}

Here we present empirical results on the total steps and wall clock time for inserting the $n + 1$th object. Recall that $n$ objects are assigned to $k$ bins in a large array. We define two metrics - total number of steps and the total wall clock time. A step is defined as one index search in the array regardless of whether or not a bin exists at that index.  

For inserting the $n + 1$th object, RJ-CH achieves as much as 20x speedup in total steps. For wall clock time, RJ-CH attains between a 2x and 7x speedup for smaller values of $\epsilon$. The speedup in wall clock time is attributed to the fewer number of full bins, practical considerations of hashing, and the cascaded overflow of CH-BL.


\begin{figure}[h!]

    \begin{minipage}[c]{0.45\linewidth}
    
    \includegraphics[width=\linewidth]{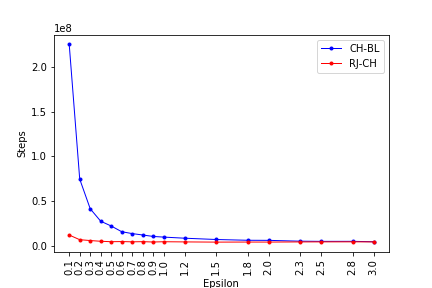}
    \caption{Total steps for adding $n + 1$th object. Note that y values are scaled.}
      \label{fig:steps}
    \end{minipage}
    \hfill
    \begin{minipage}[c]{0.45\linewidth}
    
    \includegraphics[width=\linewidth]{mainManuscriptGraphs/Time10000.png}
    \caption{Wall clock time for adding $n + 1$th object.}
      \label{fig:time2}
    \end{minipage}

\end{figure}

\section{Implementation details} \label{implementationDetails}
We use a roughly 1000 sparse array of size $2^{20}$ and the hash function Murmurhash. For each trial, we generate a pseudo-random initial string to represent each object and bin. For RJ-CH, whenever an object or bin is hashed into the array, we initialize a new counter with value 0 which is incremented until the object or bin is placed. On each iteration, the counter is converted to a string and concatenated to the pseudo-random initial string as input to Murmurhash, which produces a 128 bit number. The number is split up into four 32 bit numbers to generate the next array indexes. For an array of size $2^{20}$, we use the first 20 bits of each 32 bit hash code. For CH-BL, we generate the initial array index using Murmurhash and increment the index to walk along the array. For wall clock time and total steps, we map the array to an array of size $2^{32}$ before measuring object insertion. A single thread is used for simulations. We note that there may be benefits of running in parallel.

\section{Discussion: Biased hash functions} \label{biasedHashFunctions}
In practice, hash functions do not hash inputs to all possible outputs with equal probability. We can imagine a very biased hash function that outputs a number with 90\% probability and all other possible outputs with equal probability. This would clearly be very damaging to CH-BL. After the closest bin in the clockwise direction is full, the cascading effect will be severe. On the other hand, for RJ-CH if a bin exists at that index, after it is full RJ-CH recovers the uniform distribution of assigning objects to bins. In practice, hash functions are not nearly that biased, but we still make the below general observations about robustness against biased hash functions:

1. After the bins in the biased regions of the array are full, RJ-CH recovers the uniform distribution. CH-BL continues to worsen with the cascading effect.

2. Given that no bins are hashed into the biased regions of the array, RJ-CH recovers the uniform distribution. For CH-BL, the closest clockwise bin is severely affected.

\section{Tabulated simulation results}

Simulation results in the main manuscript in tabulated form in Tables \ref{variancetable}, \ref{searchestable}, \ref{objectstable}, \ref{binsfulltable}. A virtual bin is a virtual copy of a bin that is a reference to the bin but is in a different index in the array. 

\begin{table}[h!]
\captionof{table}{Mean and standard deviation of variance of bin loads.}
\begin{center}
\begin{tabular}{ |c|c|c|c|c|c|c|c| } 
 \hline
 objects & bins & epsilon & virtual bins & CH-BL mean & std & RJ-CH mean & std\\ 
 \hline
 10000 & 1000 & 0.1 & 0 & 6.8 & 0.2 & \textbf{2.6} & 0.1 \\ 
 10000 & 1000 & 0.3 & 0 & 19.1 & 0.4 & \textbf{6.6} & 0.2 \\ 
 10000 & 1000 & 1 & 0 & 51.9 & 1.2 & \textbf{10.0} & 0.4 \\
 10000 & 1000 & 3 & 0 & 95.0 & 3.6 & \textbf{10.0} & 0.5 \\ 
 10000 & 1000 & 0.1 & log(k) & 3.6 & 0.14 & \textbf{2.6} & 0.10 \\ 
 10000 & 1000 & 0.3 & log(k) & 10.0 & 0.29 & \textbf{6.6} & 0.22 \\ 
 10000 & 1000 & 1 & log(k) & 21.4 & 0.77 & \textbf{10.0} & 0.44 \\
 10000 & 1000 & 3 & log(k) & 24.2 & 1.16 & \textbf{10.0} & 0.46 \\ 
 3000 & 1000 & 0.1 & 0 & 2.1 & 0.04 & \textbf{1.3} & 0.04 \\ 
 3000 & 1000 & 0.3 & 0 & 2.1 & 0.04 & \textbf{1.3} & 0.04 \\ 
 3000 & 1000 & 1 & 0 & 5.3 & 0.11 & \textbf{2.6} & 0.09 \\
 3000 & 1000 & 3 & 0 & 10.0 & 0.36 & \textbf{3.0} & 0.13 \\ 
 3000 & 1000 & 0.1 & log(k) & 1.5 & 0.04 & \textbf{1.3} & 0.04 \\ 
 3000 & 1000 & 0.3 & log(k) & 1.5 & 0.04 & \textbf{1.3} & 0.04 \\ 
 3000 & 1000 & 1 & log(k) & 3.2 & 0.10 & \textbf{2.6} & 0.09 \\
 3000 & 1000 & 3 & log(k) & 4.3 & 0.20 & \textbf{3.0} & 0.13 \\ 
 \hline
\end{tabular}
\end{center}

\label{variancetable}
\end{table}

\begin{table}[h!]
\captionof{table}{Mean and standard deviation of bin searches for the $n + 1$th object to be placed.}

\begin{center}
\begin{tabular}{ |c|c|c|c|c|c|c|c| } 
 \hline
 objects & bins & epsilon & virtual bins & CH-BL mean & std & RJ-CH mean & std \\ 
 \hline
 10000 & 1000 & 0.1 & 0 & 51.52 & 68.01 & \textbf{2.79} & 2.26 \\ 
 10000 & 1000 & 0.3 & 0 & 9.31 & 11.34 & \textbf{1.31} & 0.65 \\ 
 10000 & 1000 & 1 & 0 & 2.19 & 1.76 & \textbf{1.01} & 0.09 \\
 10000 & 1000 & 3 & 0 & 1.12 & 0.38 & \textbf{1.00} & 0.00 \\ 
 10000 & 1000 & 0.1 & log(k) & 4.00 & 3.44 & \textbf{2.66} & 2.21 \\ 
 10000 & 1000 & 0.3 & log(k) & 1.82 & 1.28 & \textbf{1.33} & 0.68 \\
 10000 & 1000 & 1 & log(k) & 1.08 & 0.30 & \textbf{1.00} & 0.04 \\
 10000 & 1000 & 3 & log(k) & \textbf{1.00} & 0.03 & \textbf{1.00} & 0.00 \\ 
 3000 & 1000 & 0.1 & 0 & 10.34 & 14.06 & \textbf{1.95} & 1.36 \\ 
 3000 & 1000 & 0.3 & 0 & 9.48 & 11.85 & \textbf{1.90} & 1.30 \\ 
 3000 & 1000 & 1 & 0 & 2.35 & 2.13 & \textbf{1.08} & 0.30 \\
 3000 & 1000 & 3 & 0 & 1.17 & 0.46 & \textbf{1.00} & 0.00 \\ 
 3000 & 1000 & 0.1 & log(k) & 2.26 & 1.74 & \textbf{1.88} & 1.29 \\ 
 3000 & 1000 & 0.3 & log(k) & 2.32 & 1.71 & \textbf{1.90} & 1.31 \\ 
 3000 & 1000 & 1 & log(k) & 1.24 & 0.52 & \textbf{1.10} & 0.33 \\
 3000 & 1000 & 3 & log(k) & \textbf{1.00} & 0.05 & \textbf{1.00} & 0.00 \\ 
 \hline
\end{tabular}
\end{center}
\label{searchestable}
\end{table}

\begin{table}[h!]
\captionof{table}{Mean and standard deviation of objects placed until one bin is full.}

\begin{center}
\begin{tabular}{ |c|c|c|c|c|c|c|c| } 
 \hline
 objects & bins & epsilon & virtual bins & CH-BL mean & std & RJ-CH mean & std \\ 
 \hline
 10000 & 1000 & 0.1 & 0 & 1062 & 230 & \textbf{3295} & 477 \\ 
 10000 & 1000 & 0.3 & 0 & 1335 & 227 & \textbf{4392} & 579 \\ 
 10000 & 1000 & 1 & 0 & 2277 & 410 & \textbf{8606} & 852 \\
 10000 & 1000 & 3 & 0 & 4945 & 832 & \textbf{10000} & nan \\ 
 10000 & 1000 & 0.1 & log(k) & 2342 & 389 & \textbf{3303} & 495 \\ 
 10000 & 1000 & 0.3 & log(k) & 3027 & 447 & \textbf{4371} & 557 \\ 
 10000 & 1000 & 1 & log(k) & 5480 & 724 & \textbf{8638} & 828 \\
 10000 & 1000 & 3 & log(k) & \textbf{10000} & nan & \textbf{10000} & nan \\ 
 3000 & 1000 & 0.1 & 0 & 194 & 63 & \textbf{388} & 117 \\ 
 3000 & 1000 & 0.3 & 0 & 197 & 63 & \textbf{387} & 116 \\ 
 3000 & 1000 & 1 & 0 & 422 & 112 & \textbf{1011} & 227 \\
 3000 & 1000 & 3 & 0 & 1206 & 238 & \textbf{3000} & nan \\ 
 3000 & 1000 & 0.1 & log(k) & 331 & 104 & \textbf{378} & 116 \\ 
 3000 & 1000 & 0.3 & log(k) & 331 & 102 & \textbf{395} & 113 \\ 
 3000 & 1000 & 1 & log(k) & 816 & 186 & \textbf{1006} & 224 \\
 3000 & 1000 & 3 & log(k) & \textbf{3000} & nan & \textbf{3000} & nan \\ 
 \hline
\end{tabular}
\end{center}
\label{objectstable}
\end{table}

\begin{table}[h!]
\captionof{table}{Mean and standard deviation of percentage of bins full.}

\begin{center}
\begin{tabular}{ |c|c|c|c|c|c|c|c| } 
 \hline
 objects & bins & epsilon & virtual bins & CH-BL mean & std & RJ-CH mean & std \\ 
 \hline
 10000 & 1000 & 0.1 & 0 & 0.837 & 0.006 & \textbf{0.626} & 0.010 \\ 
 10000 & 1000 & 0.3 & 0 & 0.602 & 0.009 & \textbf{0.250} & 0.010 \\ 
 10000 & 1000 & 1 & 0 & 0.224 & 0.009 & \textbf{0.003} & 0.002 \\
 10000 & 1000 & 3 & 0 & 0.024 & 0.004 & \textbf{0.000} & 0.000 \\ 
 10000 & 1000 & 0.1 & log(k) & 0.699 & 0.009 & \textbf{0.626} & 0.009 \\ 
 10000 & 1000 & 0.3 & log(k) & 0.377 & 0.010 & \textbf{0.249} & 0.010 \\ 
 10000 & 1000 & 1 & log(k) & 0.046 & 0.006 & \textbf{0.003} & 0.002 \\
 10000 & 1000 & 3 & log(k) & \textbf{0.000} & 0.000 & \textbf{0.000} & 0.000 \\ 
 3000 & 1000 & 0.1 & 0 & 0.622 & 0.008 & \textbf{0.472} & 0.010 \\ 
 3000 & 1000 & 0.3 & 0 & 0.622 & 0.008 & \textbf{0.473} & 0.009 \\ 
 3000 & 1000 & 1 & 0 & 0.271 & 0.009 & \textbf{0.089} & 0.008 \\
 3000 & 1000 & 3 & 0 & 0.035 & 0.005 & \textbf{0.000} & 0.000 \\ 
 3000 & 1000 & 0.1 & log(k) & 0.506 & 0.009 & \textbf{0.473} & 0.009 \\ 
 3000 & 1000 & 0.3 & log(k) & 0.507 & 0.009 & \textbf{0.473} & 0.009 \\ 
 3000 & 1000 & 1 & log(k) & 0.133 & 0.008 & \textbf{0.089} & 0.007 \\
 3000 & 1000 & 3 & log(k) & 0.001 & 0.001 & \textbf{0.000} &  0.000 \\ 
 \hline
\end{tabular}
\end{center}
\label{binsfulltable}
\end{table}

\clearpage

\section{Tabulated simulation results for dynamic simulation}

We performed the same simulations with dynamic objects and bins to compare CH-BL with RJ-CH. After placing all $n$ objects, objects and bins arrive and leave at a rate of $n/k$ of objects to bins. We then observe the load balance metrics after $n$ objects have arrived or left. Results given in Tables \ref{variancetabledynamic}, \ref{searchestabledynamic}, \ref{binsfulltabledynamic}. A virtual bin is a virtual copy of a bin that is a reference to the bin but is in a different index in the array. 

\begin{table}[h!]
\captionof{table}{Mean and standard deviation of variance of bin loads.}
\begin{center}
\begin{tabular}{ |c|c|c|c|c|c|c|c|c| } 
 \hline
 objects & bins & rate & epsilon & virtual bins & CH-BL mean & std & RJ-CH mean & std \\ 
 \hline
 10000 & 1000 & ${n}/{k}$ & 0.1 & 0 & 7.1 & 1.92 & \textbf{2.6} & 0.77 \\ 
 10000 & 1000 & ${n}/{k}$ & 0.3 & 0 & 19.1 & 1.32 & \textbf{6.6} & 0.39 \\ 
 10000 & 1000 & ${n}/{k}$ & 1 & 0 & 51.9 & 1.39 & \textbf{10.0} & 0.52 \\
 10000 & 1000 & ${n}/{k}$ & 3 & 0 & 95.1 & 6.26 & \textbf{10.0} & 0.56 \\ 
 10000 & 1000 & ${n}/{k}$ & 0.1 & log(k) & 4.0 & 0.94 & \textbf{2.63} & 0.82 \\ 
 10000 & 1000 & ${n}/{k}$ & 0.3 & log(k) & 10.1 & 0.61 & \textbf{6.51} & 0.43 \\ 
 10000 & 1000 & ${n}/{k}$ & 1 & log(k) & 21.4 & 1.00 & \textbf{9.89} & 0.53 \\
 10000 & 1000 & ${n}/{k}$ & 3 & log(k) & 24.3 & 1.66 & \textbf{10.07} & 0.47 \\ 
 \hline
\end{tabular}
\end{center}
\label{variancetabledynamic}
\end{table}

\begin{table}[h!]
\captionof{table}{Mean and standard deviation of bin searches for the $n + 1$th object to be placed.}

\begin{center}
\begin{tabular}{ |c|c|c|c|c|c|c|c|c| } 
 \hline
 objects & bins & rate & epsilon & virtual bins & CH-BL mean & std & RJ-CH mean & std \\ 
 \hline
 10000 & 1000 & ${n}/{k}$ & 0.1 & 0 & 60.05 & 102.57 & \textbf{2.44} & 2.09 \\ 
 10000 & 1000 & ${n}/{k}$ & 0.3 & 0 & 10.45 & 11.94 & \textbf{1.31} & 0.59 \\ 
 10000 & 1000 & ${n}/{k}$ & 1 & 0 & 2.17 & 1.65 & \textbf{1.00} & 0.00 \\
 10000 & 1000 & ${n}/{k}$ & 3 & 0 & 1.23 & 0.53 & \textbf{1.00} & 0.00 \\ 
 10000 & 1000 & ${n}/{k}$ & 0.1 & log(k) & 4.17 & 5.26 & \textbf{2.96} & 2.89 \\ 
 10000 & 1000 & ${n}/{k}$ & 0.3 & log(k) & 1.87 & 1.55 & \textbf{1.32} & 0.68 \\ 
 10000 & 1000 & ${n}/{k}$ & 1 & log(k) & 1.07 & 0.32 & \textbf{1.01} & 0.99 \\
 10000 & 1000 & ${n}/{k}$ & 3 & log(k) & \textbf{1.00} & 0.00 & \textbf{1.00} & 0 \\ 
 \hline
\end{tabular}
\end{center}
\label{searchestabledynamic}
\end{table}

\begin{table}[h!]
\captionof{table}{Mean and standard deviation of percentage of bins full.}
\begin{center}
\begin{tabular}{ |c|c|c|c|c|c|c|c|c| } 
 \hline
 objects & bins & rate & epsilon & virtual bins & CH-BL mean & std & RJ-CH mean & std \\ 
 \hline
 10000 & 1000 & ${n}/{k}$ & 0.1 & 0 & 0.828 & 0.050 & \textbf{0.626} & 0.099 \\ 
 10000 & 1000 & ${n}/{k}$ & 0.3 & 0 & 0.601 & 0.041 & \textbf{0.249} & 0.046 \\ 
 10000 & 1000 & ${n}/{k}$ & 1 & 0 & 0.224 & 0.021 & \textbf{0.004} & 0.002 \\
 10000 & 1000 & ${n}/{k}$ & 3 & 0 & 0.025 & 0.005 & \textbf{0.000} & 0.000 \\ 
 10000 & 1000 & ${n}/{k}$ & 0.1 & log(k) & 0.688 & 0.072 & \textbf{0.630} & 0.106 \\ 
 10000 & 1000 & ${n}/{k}$ & 0.3 & log(k) & 0.378 & 0.045 & \textbf{0.250} & 0.047 \\ 
 10000 & 1000 & ${n}/{k}$ & 1 & log(k) & 0.046 & 0.011 & \textbf{0.004} & 0.003 \\
 10000 & 1000 & ${n}/{k}$ & 3 & log(k) & \textbf{0.000} & 0.000 & \textbf{0.000} & 0.000 \\ 
 \hline
\end{tabular}
\end{center}
\label{binsfulltabledynamic}
\end{table}

\clearpage

\section{Simulation comparison between Consistent Hashing with Bounded Loads with and without re-hashing on every full bin}
Here we provide empirical results showing the similar performance between CH-BL and the direct extension of CH-BL where objects are re-hashing upon encountering a full bin. Note that for objects placed until first full bin both methods are equivalent as re-hashing does not occur until there is a full bin. For the other measures, the two methods are comparable.

\begin{table}[h!]
\captionof{table}{Mean and standard deviation of variance of bin loads.}
\begin{center}
\begin{tabular}{ |c|c|c|c|c|c|c|c| } 
 \hline
 objects & bins & epsilon & virtual bins & CH-BL mean & std & re-hashing mean & std\\ 
 \hline
 10000 & 1000 & 0.1 & 0 & 6.8 & 0.2 & 6.7 & 0.2 \\ 
 10000 & 1000 & 0.3 & 0 & 19.1 & 0.4 & 19.2 & 0.4 \\ 
 10000 & 1000 & 1 & 0 & 51.9 & 1.2 & 52.1 & 1.1 \\
 10000 & 1000 & 3 & 0 & 95.0 & 3.6 & 95.1 & 3.7 \\ 
 10000 & 1000 & 0.1 & log(k) & 3.6 & 0.14 & 3.8 & 0.15 \\ 
 10000 & 1000 & 0.3 & log(k) & 10.0 & 0.29 & 10.1 & 0.29 \\ 
 10000 & 1000 & 1 & log(k) & 21.4 & 0.77 & 21.2 & 0.78 \\
 10000 & 1000 & 3 & log(k) & 24.2 & 1.16 & 24.2 & 1.17 \\ 
 \hline
\end{tabular}
\end{center}

\label{variancetablebadcomp}
\end{table}

\begin{table}[h!]
\captionof{table}{Mean and standard deviation of bin searches for the $n + 1$th object to be placed.}
\begin{center}
\begin{tabular}{ |c|c|c|c|c|c|c|c| } 
 \hline
 objects & bins & epsilon & virtual bins & CH-BL mean & std & re-hashing mean & std \\ 
 \hline
 10000 & 1000 & 0.1 & 0 & 51.52 & 68.01 & 70.55 & 74.24 \\ 
 10000 & 1000 & 0.3 & 0 & 9.31 & 11.34 & 9.19 & 8.71 \\ 
 10000 & 1000 & 1 & 0 & 2.19 & 1.76 & 2.21 & 1.57 \\
 10000 & 1000 & 3 & 0 & 1.12 & 0.38 & 1.13 & 0.38 \\ 
 10000 & 1000 & 0.1 & log(k) & 4.00 & 3.44 & 5.28 & 4.58 \\ 
 10000 & 1000 & 0.3 & log(k) & 1.82 & 1.28 & 2.08 & 1.52 \\
 10000 & 1000 & 1 & log(k) & 1.08 & 0.30 & 1.09 & 0.29 \\
 10000 & 1000 & 3 & log(k) & 1.00 & 0.03 & 1.00 & 0.00 \\ 
 \hline
\end{tabular}
\end{center}
\label{searchestablebadcomp}
\end{table}

\begin{table}[h!]
\captionof{table}{Mean and standard deviation of percentage of bins full.}
\begin{center}
\begin{tabular}{ |c|c|c|c|c|c|c|c| } 
 \hline
 objects & bins & epsilon & virtual bins & CH-BL mean & std & re-hashing mean & std \\ 
 \hline
 10000 & 1000 & 0.1 & 0 & 0.837 & 0.006 & 0.836 & 0.006 \\ 
 10000 & 1000 & 0.3 & 0 & 0.602 & 0.009 & 0.603 & 0.009 \\ 
 10000 & 1000 & 1 & 0 & 0.224 & 0.009 & 0.223 & 0.009 \\
 10000 & 1000 & 3 & 0 & 0.024 & 0.004 & 0.024 & 0.004 \\ 
 10000 & 1000 & 0.1 & log(k) & 0.699 & 0.009 & 0.714 & 0.009 \\ 
 10000 & 1000 & 0.3 & log(k) & 0.377 & 0.010 & 0.389 & 0.009 \\ 
 10000 & 1000 & 1 & log(k) & 0.046 & 0.006 & 0.046 & 0.006 \\
 10000 & 1000 & 3 & log(k) & 0.000 & 0.000 & 0.000 & 0.000 \\ 
 \hline
\end{tabular}
\end{center}
\label{binsfulltablebadcomp}
\end{table}

\begin{table}[h!]
\captionof{table}{Mean and standard deviation of total steps given by $10^6$ multiplied by the shown value. }
\begin{center}
\begin{tabular}{ |c|c|c|c|c|c|c|c| } 
 \hline
 objects & bins & epsilon & virtual bins & CH-BL mean & std & re-hashing mean & std\\ 
 \hline
 10000 & 1000 & 0.1 & 0 & 230  & 300  & 296  & 300  \\ 
 10000 & 1000 & 0.3 & 0 & 41  & 60  & 44  & 40  \\ 
 10000 & 1000 & 1 & 0 & 9.7  & 10  & 9.6  & 10  \\
 10000 & 1000 & 3 & 0 & 4.4  & 5.0  & 4.7  & 6.0  \\ 
 10000 & 1000 & 0.1 & log(k) & 2.4  & 3.0  & 3.4  & 4.0  \\ 
 10000 & 1000 & 0.3 & log(k) & 1.1  & 1.0  & 1.2  & 1.0  \\ 
 10000 & 1000 & 1 & log(k) & 0.7  & 0.7  & 0.7  & 0.7 \\
 10000 & 1000 & 3 & log(k) & 0.6  & 0.6  & 0.6  & 0.7  \\ 
 \hline
\end{tabular}
\end{center}
\label{totalstepstablebadcomp}
\end{table}

\clearpage

\section{Comparison for virtual bins and supplementary figures}
\label{SupplementaryFigures}

We generate $n$ objects and $k$ bins with $v$ virtual copies of each bin where each bin has capacity $C = \ceiling{\frac{n}{k}(1 + \epsilon)} $. A virtual copy of a bin is a reference to the bin that is in a different index in the array. Virtual bins may be used to improve wall clock time. We hash each of the $kv$ bins into a large array with few collisions. We then populate the bins according to the two methods of RJ-CH and CH-BL. We resolve bin collisions by rehashing. We note here that virtual copies can be undesirable due to bin collisions, especially when the total number of bins is large. 

We use the pairs (1000 objects, 1000 bins), and (3000 objects, 1000 bins). For each pair, we use no virtual bins or $log(k)$ virtual bins. 
We try all $\epsilon \in \{0.1, 0.2, 0.3, 0.4, 0.5, 0.6, 0.7, 0.8, 0.9, 1, 1.2, 1.5, 1.8, 2.0, 2.3, 2.5, 2.8, 3\}$, and perform 1000 trials for each pair and $\epsilon$ where we initialize each trial from scratch.

\begin{figure}[h!]
  \centering
  \begin{subfigure}[b]{0.4\linewidth}
    \includegraphics[width=\linewidth]{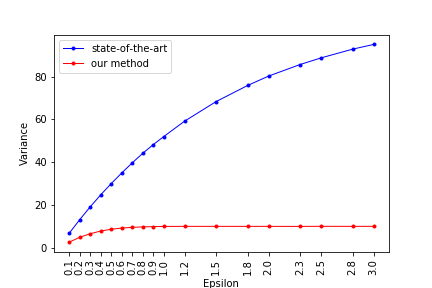}
    \caption{10000 objects, 1000 bins, no virtual bins. \\}
  \end{subfigure}
  \begin{subfigure}[b]{0.4\linewidth}
    \includegraphics[width=\linewidth]{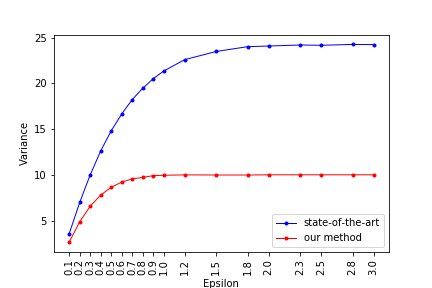}
    \caption{10000 objects, 1000 bins, log(k) virtual bins.}
  \end{subfigure}
    \begin{subfigure}[b]{0.4\linewidth}
    \includegraphics[width=\linewidth]{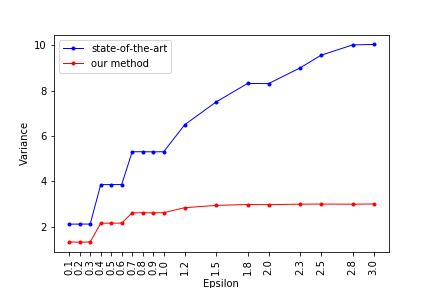}
    \caption{3000 objects, 1000 bins, no virtual bins. \\ \: \:}
  \end{subfigure}
  \begin{subfigure}[b]{0.4\linewidth}
    \includegraphics[width=\linewidth]{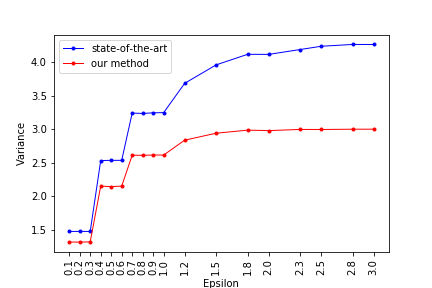}
    \caption{3000 objects, 1000 bins, log(k) virtual bins.}
  \end{subfigure}
  \caption{Variance of bin loads for different configurations.}
  \label{fig:varianceSupplementary}
\end{figure}

\begin{figure}[h!]
  \centering
  \begin{subfigure}[b]{0.4\linewidth}
    \includegraphics[width=\linewidth]{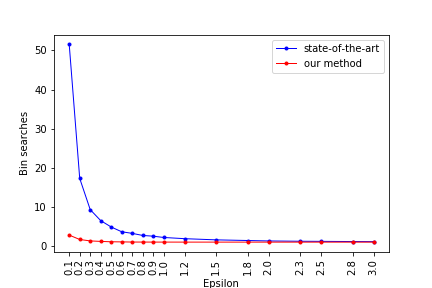}
    \caption{10000 objects, 1000 bins, no virtual bins. \\ \: \:}
  \end{subfigure}
  \begin{subfigure}[b]{0.4\linewidth}
    \includegraphics[width=\linewidth]{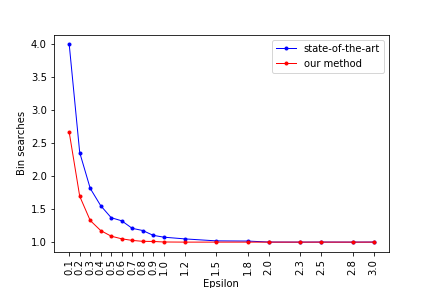}
    \caption{10000 objects, 1000 bins, log(k) virtual bins.}
  \end{subfigure}
    \begin{subfigure}[b]{0.4\linewidth}
    \includegraphics[width=\linewidth]{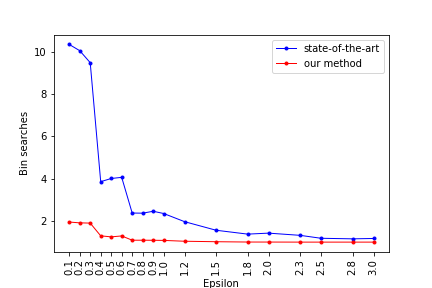}
    \caption{3000 objects, 1000 bins, no virtual bins.\\ \: \:}
  \end{subfigure}
  \begin{subfigure}[b]{0.4\linewidth}
    \includegraphics[width=\linewidth]{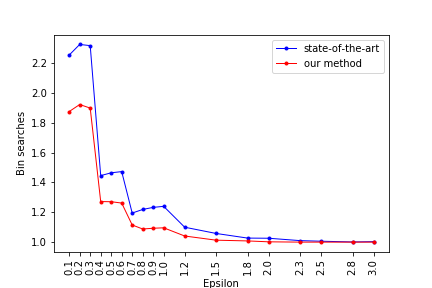}
    \caption{3000 objects, 1000 bins, log(k) virtual bins.}
  \end{subfigure}
  \caption{Bin searches for the $n + 1$th object to be placed for different configurations.}
  \label{fig:searchesSupplementary}
\end{figure}

\begin{figure}[h!]
  \centering
  \begin{subfigure}[b]{0.4\linewidth}
    \includegraphics[width=\linewidth]{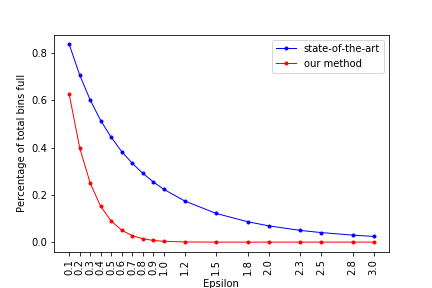}
    \caption{10000 objects, 1000 bins, no virtual bins.\\ \: \:}
  \end{subfigure}
  \begin{subfigure}[b]{0.4\linewidth}
    \includegraphics[width=\linewidth]{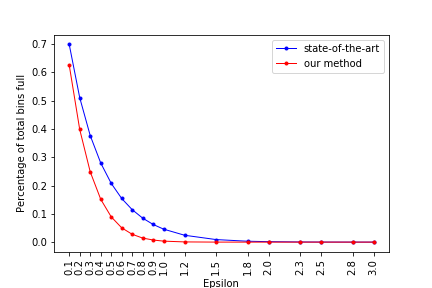}
    \caption{10000 objects, 1000 bins, log(k) virtual bins.}
  \end{subfigure}
    \begin{subfigure}[b]{0.4\linewidth}
    \includegraphics[width=\linewidth]{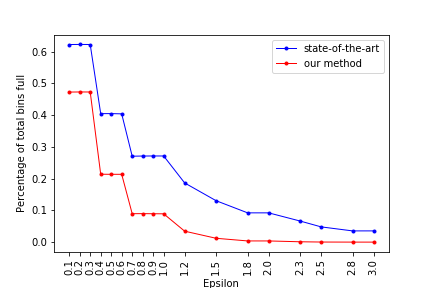}
    \caption{3000 objects, 1000 bins, no virtual bins.\\ \: \:}
  \end{subfigure}
  \begin{subfigure}[b]{0.4\linewidth}
    \includegraphics[width=\linewidth]{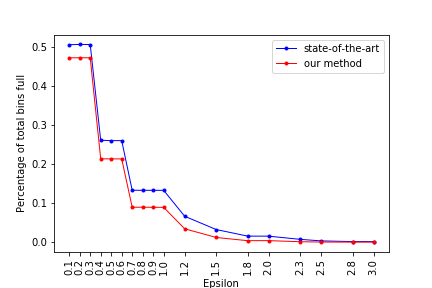}
    \caption{3000 objects, 1000 bins, log(k) virtual bins.}
  \end{subfigure}
  \caption{Percentage of total bins full for different configurations.}
  \label{fig:binsfullSupplementary}
\end{figure}

\begin{figure}[h!]
  \centering
  \begin{subfigure}[b]{0.4\linewidth}
    \includegraphics[width=\linewidth]{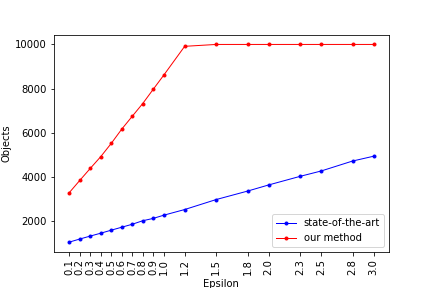}
    \caption{10000 objects, 1000 bins, no virtual bins.\\ \: \:}
  \end{subfigure}
  \begin{subfigure}[b]{0.4\linewidth}
    \includegraphics[width=\linewidth]{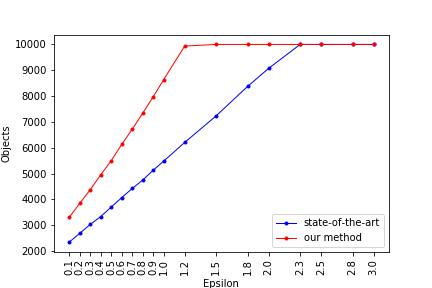}
    \caption{10000 objects, 1000 bins, log(k) virtual bins.}
  \end{subfigure}
    \begin{subfigure}[b]{0.4\linewidth}
    \includegraphics[width=\linewidth]{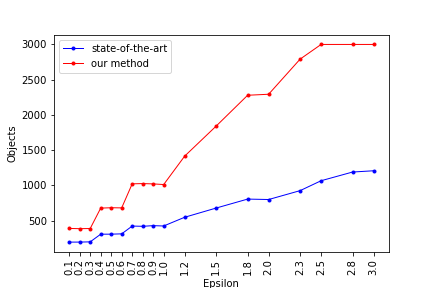}
    \caption{3000 objects, 1000 bins, no virtual bins.\\ \: \:}
  \end{subfigure}
  \begin{subfigure}[b]{0.4\linewidth}
    \includegraphics[width=\linewidth]{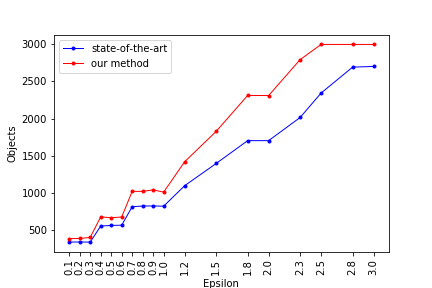}
    \caption{3000 objects, 1000 bins, log(k) virtual bins.}
  \end{subfigure}
  \caption{Number of objects placed until one bin is full.}
  \label{fig:objectsTillFirstFullSupplementary}
\end{figure}

\begin{figure}[h!]
  \centering
  \begin{subfigure}[b]{0.4\linewidth}
    \includegraphics[width=\linewidth]{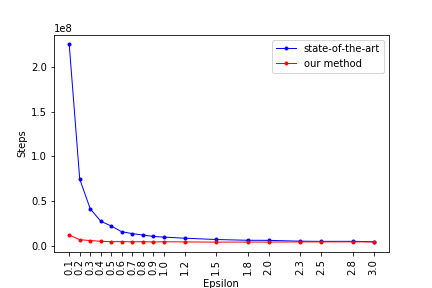}
    \caption{10000 objects, 1000 bins, no virtual bins.\\ \: \:}
  \end{subfigure}
  \begin{subfigure}[b]{0.4\linewidth}
    \includegraphics[width=\linewidth]{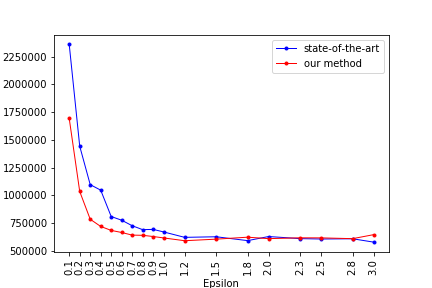}
    \caption{10000 objects, 1000 bins, log(k) virtual bins.}
  \end{subfigure}
    \begin{subfigure}[b]{0.4\linewidth}
    \includegraphics[width=\linewidth]{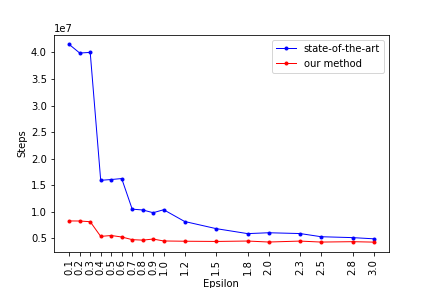}
    \caption{3000 objects, 1000 bins, no virtual bins.\\ \: \:}
  \end{subfigure}
  \begin{subfigure}[b]{0.4\linewidth}
    \includegraphics[width=\linewidth]{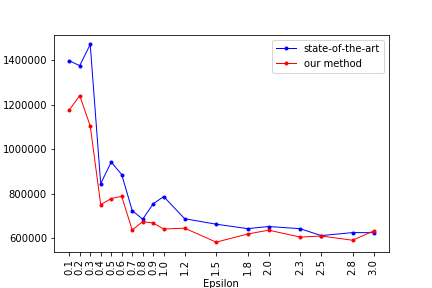}
    \caption{3000 objects, 1000 bins, log(k) virtual bins.}
  \end{subfigure}
  \caption{Total steps for adding $n + 1$th object for different configurations. Note that some y values are scaled.}
  \label{fig:stepsSupplementary}
\end{figure}

\begin{figure}[h!]
  \centering
  \begin{subfigure}[b]{0.4\linewidth}
    \includegraphics[width=\linewidth]{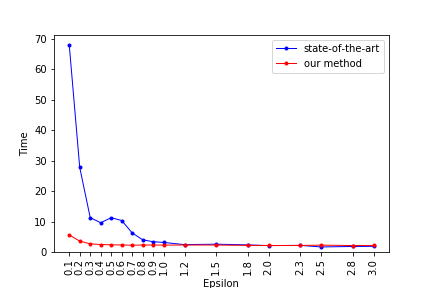}
    \caption{10000 objects, 1000 bins, no virtual bins.\\ \: \:}
  \end{subfigure}
  \begin{subfigure}[b]{0.4\linewidth}
    \includegraphics[width=\linewidth]{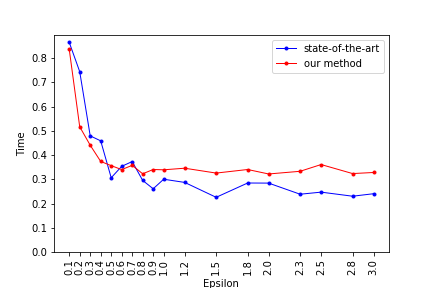}
    \caption{10000 objects, 1000 bins, log(k) virtual bins.}
  \end{subfigure}
    \begin{subfigure}[b]{0.4\linewidth}
    \includegraphics[width=\linewidth]{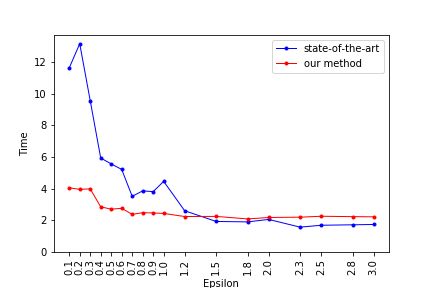}
    \caption{3000 objects, 1000 bins, no virtual bins.\\ \: \:}
  \end{subfigure}
  \begin{subfigure}[b]{0.4\linewidth}
    \includegraphics[width=\linewidth]{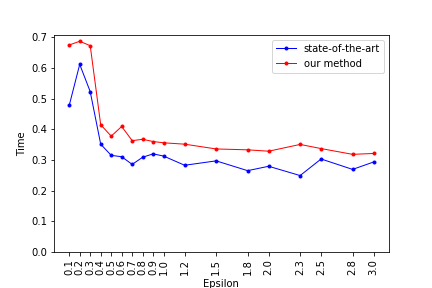}
    \caption{3000 objects, 1000 bins, log(k) virtual bins.}
  \end{subfigure}
  \caption{Wall clock time for adding $n + 1$th object for different configurations.}
  \label{fig:timeSupplementary}
\end{figure}

\begin{figure}[h!]
  \centering
  \begin{subfigure}[b]{0.4\linewidth}
    \includegraphics[width=\linewidth]{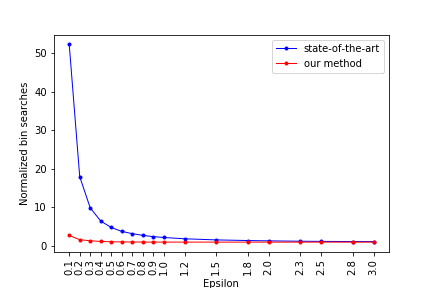}
    \caption{10000 objects, 1000 bins, no virtual bins.\\ \: \:}
  \end{subfigure}
  \begin{subfigure}[b]{0.4\linewidth}
    \includegraphics[width=\linewidth]{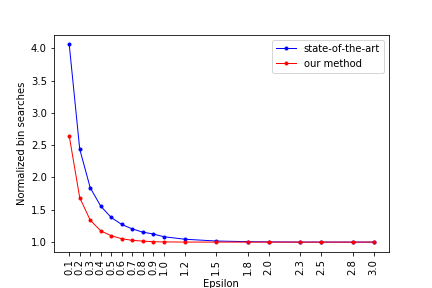}
    \caption{10000 objects, 1000 bins, log(k) virtual bins.}
  \end{subfigure}
    \begin{subfigure}[b]{0.4\linewidth}
    \includegraphics[width=\linewidth]{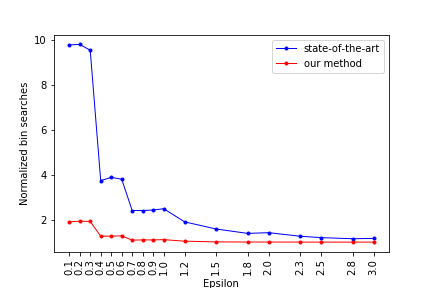}
    \caption{3000 objects, 1000 bins, no virtual bins.\\ \: \:}
  \end{subfigure}
  \begin{subfigure}[b]{0.4\linewidth}
    \includegraphics[width=\linewidth]{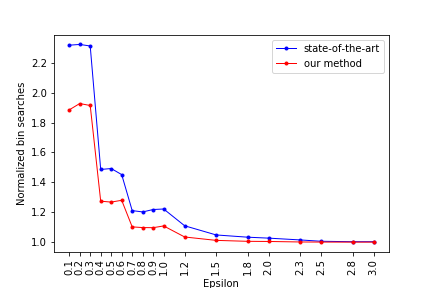}
    \caption{3000 objects, 1000 bins, log(k) virtual bins.}
  \end{subfigure}
  \caption{Normalized bin searches for bin removal to be placed for different configurations.}
  \label{fig:searchesBinRemovalSupplementary}
\end{figure}

\begin{figure}[h!]
  \centering
  \begin{subfigure}[b]{0.4\linewidth}
    \includegraphics[width=\linewidth]{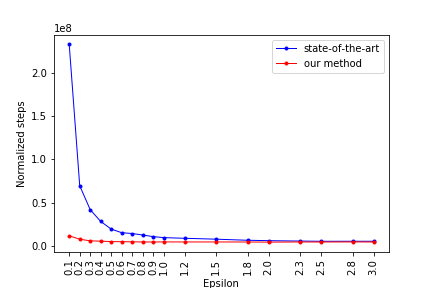}
    \caption{10000 objects, 1000 bins, no virtual bins.\\ \: \:}
  \end{subfigure}
  \begin{subfigure}[b]{0.4\linewidth}
    \includegraphics[width=\linewidth]{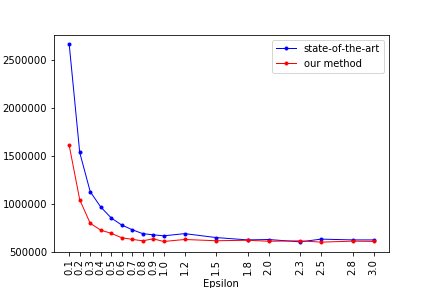}
    \caption{10000 objects, 1000 bins, log(k) virtual bins.}
  \end{subfigure}
    \begin{subfigure}[b]{0.4\linewidth}
    \includegraphics[width=\linewidth]{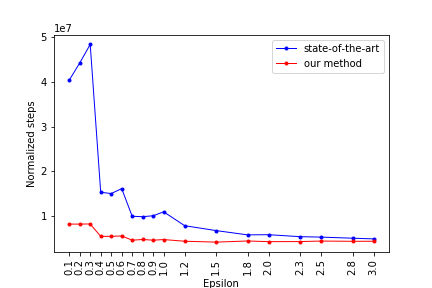}
    \caption{3000 objects, 1000 bins, no virtual bins.\\ \: \:}
  \end{subfigure}
  \begin{subfigure}[b]{0.4\linewidth}
    \includegraphics[width=\linewidth]{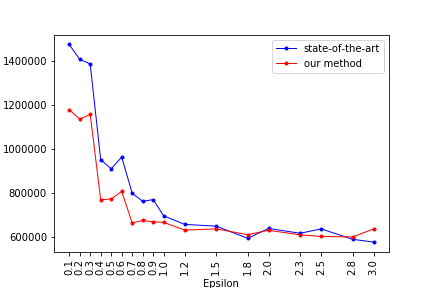}
    \caption{3000 objects, 1000 bins, log(k) virtual bins.}
  \end{subfigure}
  \caption{Per object total steps for removing a bin for different configurations. Note that some y values are scaled.}
  \label{fig:stepsBinSupplementary}
\end{figure}

\begin{figure}[h!]
  \centering
  \begin{subfigure}[b]{0.4\linewidth}
    \includegraphics[width=\linewidth]{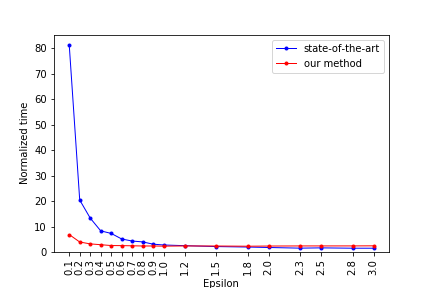}
    \caption{10000 objects, 1000 bins, no virtual bins.\\ \: \:}
  \end{subfigure}
  \begin{subfigure}[b]{0.4\linewidth}
    \includegraphics[width=\linewidth]{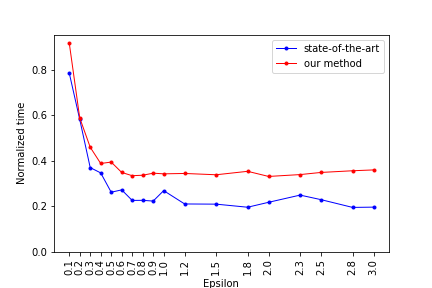}
    \caption{10000 objects, 1000 bins, log(k) virtual bins.}
  \end{subfigure}
    \begin{subfigure}[b]{0.4\linewidth}
    \includegraphics[width=\linewidth]{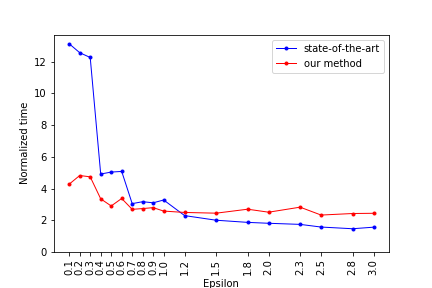}
    \caption{3000 objects, 1000 bins, no virtual bins.\\ \: \:}
  \end{subfigure}
  \begin{subfigure}[b]{0.4\linewidth}
    \includegraphics[width=\linewidth]{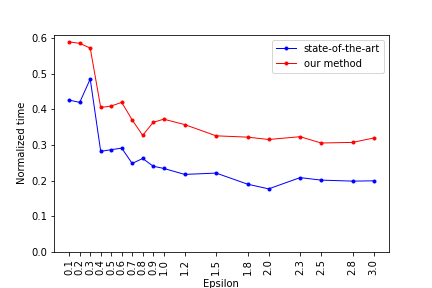}
    \caption{3000 objects, 1000 bins, log(k) virtual bins.}
  \end{subfigure}
  \caption{Per object wall clock time for removing a bin for different configurations.}
  \label{fig:timeBinSupplementary}
\end{figure}

\end{appendices}

\end{document}